\documentclass[a4paper,twocolumn,11pt,accepted=2022-09-08]{quantumarticle}
\pdfoutput=1
\usepackage[utf8]{inputenc}
\usepackage[english]{babel}
\usepackage[T1]{fontenc}
\usepackage{amsmath}
\usepackage{amssymb}
\usepackage{hyperref}
\usepackage{tikz}
\usepackage{lipsum}% only package missing from oficial template
\usepackage[numbers,sort&compress]{natbib}
\usepackage{graphicx}
\usepackage{dcolumn}
\usepackage{bm}
\usepackage{ae}
\usepackage{xfrac}  
\usepackage{physics}
\RequirePackage{dsfont}

\usepackage{xcolor}
\usepackage{comment}
\usepackage{units}
\usepackage[normalem]{ulem}

\begin{document}

\title{Enhanced Photonic Maxwell's Demon with Correlated Baths}

\author{Guilherme L. Zanin$^{1,2,\dagger}$}
\author{Michael Antesberger$^1$}
\author{Maxime J. Jacquet$^{1,3}$}
\author{Paulo H. Souto Ribeiro$^2$}
\author{Lee A. Rozema$^1$}
\author{Philip Walther$^{1,4}$}

\affiliation{$^1$University of Vienna, Faculty of Physics, Vienna Center for Quantum Science and Technology (VCQ), Vienna, Austria\\
$^2$Departamento de F\'{i}sica, Universidade Federal de Santa Catarina, Florian\'{o}polis, Santa Catarina 88040-900, Brazil\\
$^3$Laboratoire Kastler Brossel, Sorbonne Universit\'e, CNRS, ENS-Universit\'e PSL, Coll\`ege de France, Paris 75005, France.\\
$^4$Christian Doppler Laboratory for Photonic Quantum Computer, Faculty of Physics,  University of Vienna, 1090 Vienna, \\Austria\\
$^\dagger$Correspondence to:  guilherme.zanin@univie.ac.at.}
\maketitle

\begin{abstract}
Maxwell’s Demon is at the heart of the interrelation between quantum information processing and thermodynamics. In this thought experiment, 
a demon generates a temperature gradient between two thermal baths initially at equilibrium by gaining information at the single-particle level and applying classical feed-forward operations, allowing for the extraction of work.
Here we implement a photonic version of Maxwell's Demon with active feed-forward in a fibre-based system using ultrafast optical switches. 
We experimentally show that, if correlations exist between the two thermal baths, the Demon can generate a temperature difference over an order of magnitude larger than without correlations, and so extract more work. Our work demonstrates the great potential of photonic experiments -- which provide a unique degree of control on the system -- to access new regimes in quantum thermodynamics. 
\end{abstract}

\section{Introduction}
\begin{figure*}[t]
    \centering\includegraphics[width=\textwidth]{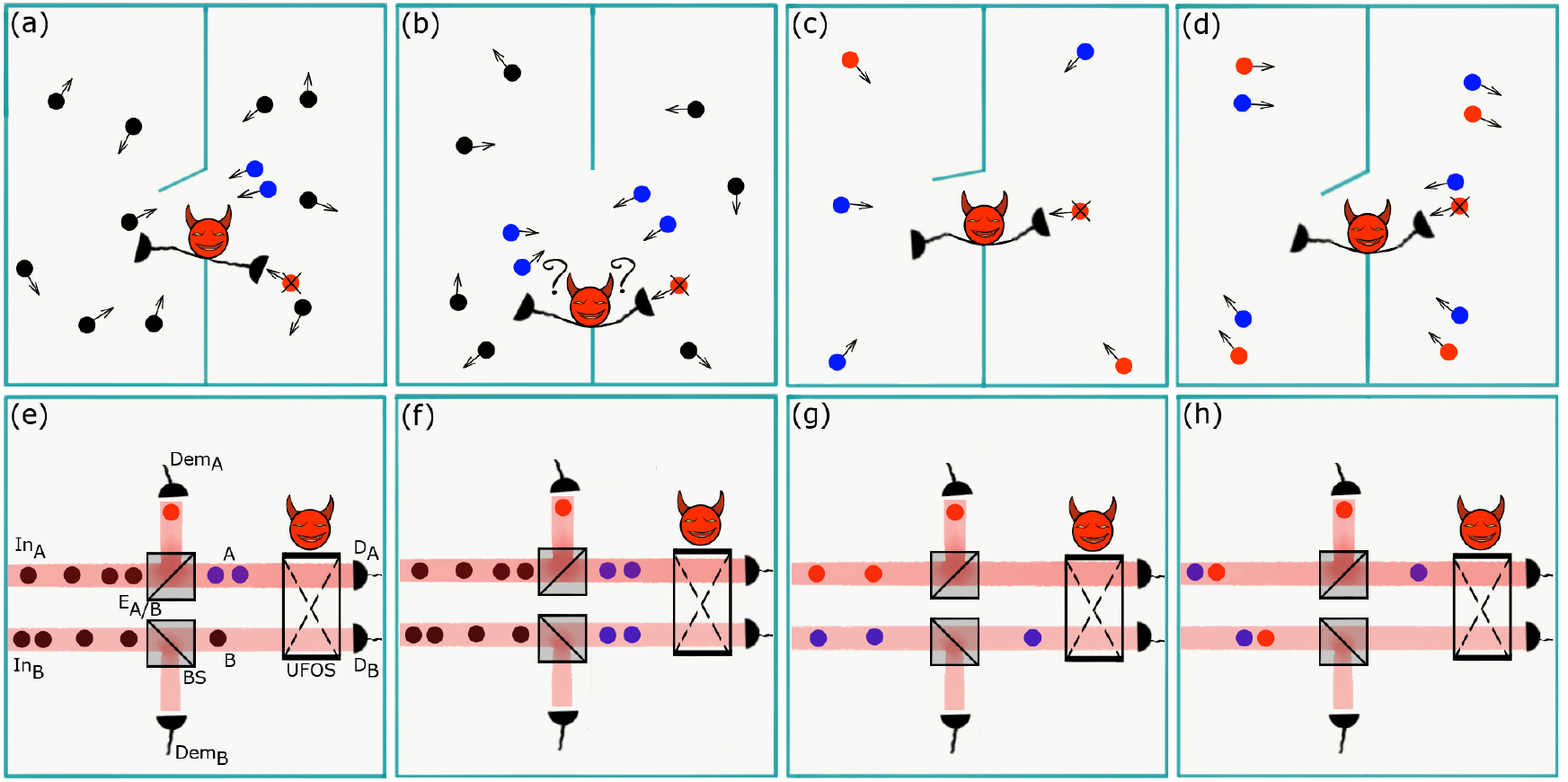}
    \caption{\textbf{Principle of operation of the photonic Maxwell's Demon (MD):} the MD controls the path of particles/photons in the system. The colours red and blue represent correlated photons; i.e. when the demon measures a red particle, it learns the location of the blue particle. Top row, particle picture of the photonic MD; bottom row, actual implementation of the photonic MD. \textbf{(a)} and \textbf{(e)}, uncorrelated thermal states; \textbf{(b)} and \textbf{(f)}, split thermal states; \textbf{(c)} and \textbf{(g)} correlated states; \textbf{(d)} and \textbf{(h)}, anti-correlated states. \textbf{(e)} The mode labels shown here also apply to panels \textbf{(f)}-\textbf{(g)}. $\mathrm{In}_{A/B}$ and $\mathrm{E}_{A/B}$ label the input modes of the MD's beamsplitter. The various thermal states are input into $\mathrm{In}_{A/B}$, while vacuum is always incident in modes  $\mathrm{E}_{A/B}$. We use $A/B$ and $\mathrm{Dem}_{A/B}$ to label the output modes of the beamsplitters.  Modes $\mathrm{Dem}_{A/B}$ are sent to the MD's detectors, while $A/B$ are sent to UFOS so the MD can swap them if it wishes. Finally, $\mathrm{D}_{A/B}$ label the output modes of the UFOS, which are directly sent to the detectors to measure the photon number imbalance.
    In the photonic implementation without correlations (\textbf{(a)} and \textbf{(e)}), a detection at the demon's detector $\mathrm{Dem}_A$ heralds more photons in mode $A$ (indicated by the two blue particles), corresponding to an increased temperature. 
    With classical correlations arising from split thermal baths (\textbf{(b)} and \textbf{(f)}), detection at $\mathrm{Dem}_A$ increases the photon number in both mode $A$ and mode $B$ by the same amount, rendering the demon powerless.
    The MD with correlated (\textbf{(c)} and \textbf{(g)}) and anti-correlated (\textbf{(d)} and \textbf{(h)}) baths learns about particles in both baths. Detecting the red photon at $\mathrm{Dem}_A$ tells the MD that there is now one more photon in mode $B$ for correlated baths. While detecting the red photon at $\mathrm{Dem}_A$ for for anti-correlated baths tells the demon that there are no photons in mode $B$.}
    \label{fig:md_mosaic}
\end{figure*}

Thermodynamics was conceived as a phenomenological theory for  the  equilibrium  properties  of  macroscopic systems ranging from gas (ensembles of `many' small systems) to black holes.
For these systems, theories like that of Maxwell's for heat~\cite{maxwell_1871} provide a complete description of quantities such as temperature or work, and the manipulation of thermal states -- as well as their emergence from fluctuations at the smallest scale -- can be described in terms of (quantum) information processing~\cite{Landauer_irrev_1961,del_rio_resource_2015,goold_role_2016}.
Reciprocally, thermodynamic considerations ought to limit quantum information processing itself~\cite{Oppenheimer_thermoquant_2002,Zurek_discord_2003,Vedral_thermoquant_2005}.
This interrelation is perhaps best manifested in the protocol known as `Maxwell’s demon' (MD)~\cite{maxwell_1871,THOMSON1874,leffMaxwellDemonEntropy1990,plenioPhysicsForgettingLandauer2001,maruyamaColloquiumPhysicsMaxwell2009,parrondoThermodynamicsInformation2015,rexMaxwellDemonHistorical2017,beyerSteeringHeatEngines2019,huMaxwellDemonQuantum2021}.
As illustrated in Fig. \ref{fig:md_mosaic} \textbf{(a)}, the MD monitors the motion of gas particles inside a partitioned box.
Based on its previous knowledge of the particle's velocity~\cite{Szilard1929,bennettThermodynamicsComputationReview1982}, the MD can sort fast particles from slow particles by opening an aperture in the partition such that all fast particles eventually end up on one side of the partition and all slow particles on the other.
So the MD brings the system out of equilibrium, and one can then exploit the engineered temperature differential to extract work.
In modern experiments~\cite{Toyabe2010_particle_MD,Berut2012_Experimental_laundaer_MD,JKoski_experimental_mutualInformation_MD,Debiossac2020_univie_MD,Maxwell_Lesser_Demon_MD,PhysRevLett.121.180601_MD,Autonomous_MD,Ultracold_MD,quantum_MD,solid-state_quantum_MD,superconducting1_MD,MD_quantum-zeno_regime}, {we say that the MD generates work by measuring individual particles to gain information~\cite{Szilard1929,bennettThermodynamicsComputationReview1982}, and depending on these measurement results, it applies different operations to the bath (i.e. opening the door, or leaving it closed).
These measurement-dependent operations are called feed-forward operations.}
Here, work and energy are bound to the available information, and are ultimately physically limited by it~\cite{bennettThermodynamicsComputationReview1982,sagawa_2ndlaw_information}.

The `power' of the MD, the temperature imbalance it can generate per measurement, can be greatly enhanced by operating with correlated particles: Consider a situation where, for every particle moving in a certain direction in the left bath, there is a twin particle moving in the same direction in the right bath (see Fig.~\ref{fig:md_mosaic} \textbf{(c)}).
If the MD observes a particle in the left bath it also learns something about an (unobserved) particle in the right bath.
Because of this additional information, the MD can  generate a larger temperature difference per measurement~\cite{Theo_shu_MD} by simply modifying its feed-forward protocol to make use of the new information.

While correlations are difficult to engineer in most systems, doing so is very easy in photonics.
Here, we present the first experimental realisation of a photonic MD operating with correlations and active feed-forward.
The photonic MD was first proposed and a proof-of-principle experiment was demonstrated in \cite{Barbieri16MD}, and the idea to enhance the MD's power was proposed in \cite{Theo_shu_MD}.

The working principle of an uncorrelated photonic MD is depicted in Fig.~\ref{fig:md_mosaic} (\textbf{e}).
In a photonic MD, the baths are realised with optical modes, each containing a thermal state of photons whose average photon number directly sets the temperature of the bath.
These two optical modes are then sent to individual detectors, $D_A$ and $D_B$.
When the two optical modes have an equal photon number {(the baths have an equal temperature)}, the detectors register the same counts and no work can be extracted.
When a photon is subtracted from this state at a beamsplitter and detected in the reflection port with detector $\mathrm{Dem}_A$ or $\mathrm{Dem}_B$, bunching will occur, meaning that the photon number in the transmitted beam will be temporarily doubled (as depicted in red in Fig.~\ref{fig:md_mosaic} (\textbf{e})).
This photon extraction provides information about the average photon number in the beam, based on which a classical feed-forward operation may then be applied to create an imbalance in the photon number between the two beams. This energy imbalance can in turn be used to extract work.
The gain of information and the generation of a temperature difference through classical feed-forward is precisely the functioning of an MD --- a photonic MD is thus a device that can generate a photon number imbalance between two thermal states that initially have the same photon number.
{The power of a photonic MD is simply quantified by the photon number difference that can be generated to extract work.}

The working principle of the photonic MD was demonstrated only recently in a proof-of-principle experiment~\cite{Barbieri16MD}.
There, the feed-forward at the core of the extraction of work was not applied, instead the operation of the MD was shown by post-processing the data.
Detectors $D_A$ and $D_B$ were replaced by photodiodes connected to a capacitor.
In the case where the photon number at $D_A$ and $D_B$ was balanced, the photo-currents generated by the photodiodes cancelled out, creating a zero net charge on the capacitor.
If, however, there were more photons in one beam, the capacitor's polarity could have been flipped (classical feed-forward operation) such that the photo-currents would no longer cancel out and the capacitor would always be charged positively, allowing the extraction of work.
In~\cite{Barbieri16MD}, instead of flipping the capacitor's polarity, the operation of the photonic MD was quantified by recording the charge across the capacitor in real time.
It was found that a positive (negative) charge of the capacitor would correspond to a click at $\mathrm{Dem}_A$ ($\mathrm{Dem}_B$).
Hence, no work was extracted.

In our photonic MD, we implement active feed-forward to physically swap the paths of the beams dependent on detection events at $\mathrm{Dem}_A$ and $\mathrm{Dem}_B$ using an ultra-fast optical switch (methods adapted from~\cite{Zanin_fibre_compatible}).
While our active feed-forward would allow for the extraction of work, we do not extract work here. We instead monitor the photon number imbalance directly using single-photon detectors, as this characterises the maximum work that could be extracted.
This detection scheme is also more compatible with the quantum states of light that we use to generate correlations to increase the power of the MD.

Correlations in quantum thermodynamics have been used in numerous experiments in quantum thermodynamics~\cite{DornerquantumWorkwithout2projectiveEnergy,RoncagliaWorkMeasurement,BatalhaoExperimentalWorkdistribuition,Shuoming_quantumJar_iontrap,landauer_experimentalPeterson,single_atom_heat_engine,Single_qubit_thermometry,zhang_wang_adiabiatic_workflutuationexp,Exp_single_trap_ionJarzynski_Xiong_Vedral,quantumThermoAdiabaticExpeirmental_hu2020,PRA_jarzynski_turbulence_giqsul}, but not in terms of information processing by the MD.
Here, we compare the statistics obtained with uncorrelated thermal states (produced classically) and split thermal states to those obtained with correlated and anticorrelated thermal states produced by spontaneous parametric down-conversion and a path-entangled (so-called N00N~\cite{dowling_noon_2002}) state, respectively.
We show that the choice of correlations is essential in enabling the MD to extract work.
Our experiments evidence that particular correlations allow for an order of magnitude increase of the power of the MD, while others prevent the MD from extracting any work at all.
Furthermore, we derive a figure of merit for the power of the MD in the regime of low photon-numbers and demonstrate that information processing with correlations and anti-correlations vastly outperforms operation with the other states.

Although we generate some of these correlations using quantum states of light, the enhancement of the power of the MD does not rely on the quantum nature of the correlations.
It remains an open question if quantum correlations could further enhance the power of the MD.
Nevertheless, correlations are ubiquitous in quantum information processing, and so our photonic implementation settles Maxwell's demon therein.

\section{The power of Maxwell's Demon}

\begin{figure*}[t]
    \centering\includegraphics[width=\textwidth]{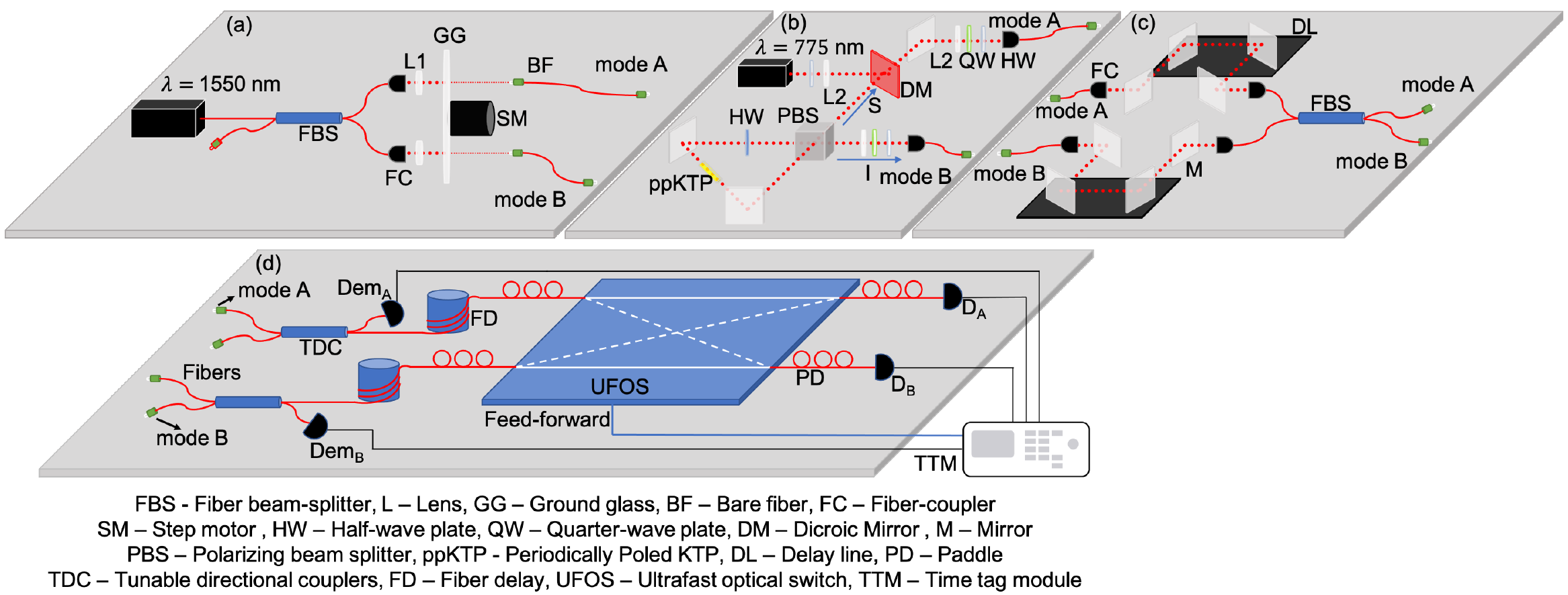}
    \caption{\textbf{Active photonic Maxwell's Demon.} The two photon sources: \textbf{(a)}, uncorrelated thermal states and \textbf{(b)}, correlated thermal states (Sagnac loop for SPDC). Splitting one mode of one of the sources shown in \textbf{(a)} on a beamsplitter (not shown) yields the split thermal state. Passing photons from \textbf{(b)} through the Hong-Ou-Mandel interferometer \textbf{(c)} yields the two-photon N00N state. The Maxwell's Demon in \textbf{(d)} monitors the statistics of the incoming photons with the detectors in modes Dem$_A$ and Dem$_B$ and conditionally controls the path the photons take through the ultra-fast optical switch (UFOS) to the detectors in modes $D_A$ and $D_B$.
}
    \label{fig:setup}
\end{figure*}

In our experiment, the MD consists of a logic that monitors photon arrival events at two detectors and conditionally controls the path light takes.
Specifically, the experiment (Fig.~\ref{fig:setup} \textbf{(d)}) is based on two separate spatial modes for light propagation, corresponding to two paths (labelled $\mathrm{\mathrm{In_A}}$ and $\mathrm{In_B}$ in Fig.~\ref{fig:md_mosaic} \textbf{(e)}) in the setup, in which we send states with various statistics.
The MD consists of two single-photon detectors ($\mathrm{Dem}_A$ and $\mathrm{Dem}_B$), two variable-reflectivity beamsplitters (implemented with tunable directional couplers, labelled TDC), the ultra-fast optical switches (UFOS), and the logic (labelled TTM) used to control the UFOS.
The detectors monitor photon arrival events in the reflected port of the TDCs, and send signals to the coincidence logic which then sets the state of the UFOS to route the light. 

The feed-forward protocol functions as follows: The thermal states (generated by either of the sources Fig.~\ref{fig:setup} \textbf{(a)}, \textbf{(b)} or \textbf{(b)}+\textbf{(c)}) enter the setup (Fig.~\ref{fig:setup}  \textbf{(d)}) in modes {$\mathrm{In_A}$ and $\mathrm{In_B}$}.
First, the temperature of the thermal baths are set equal by fixing the UFOS in the `bar state' (to transmit mode $A$ to $D_A$ and $B$ to $D_B$), and equalising the photon rates $N_A$ and $N_B$ measured by single-photon detectors in modes $D_A$ and $D_B$.
Second, we make the MD's feedforward active.
Microscopic knowledge about incoming photons in modes $\mathrm{In_A}$ and $\mathrm{In_B}$ is acquired via the detector clicks in the reflected arms $\mathrm{Dem}_A$ and $\mathrm{Dem}_B$.
Based on this information, the MD operates the UFOS (or not) in order to route the beam with a higher photon number in mode $A$ or $B$ to $D_A$, thus creating a photon-number imbalance at the output.
The exact details of this routing depend on the states considered and we discuss them below.

To characterise the temperature gradient generated by the MD, we directly measure the photon rates $N_A$ and $N_B$ in the two beams exiting the experiment with single-photon detectors in modes $D_A$ and $D_B$ when the feed-forward is active.
We then measure $\Delta N=N_A-N_B$, the photon number difference between these detectors.
Although we do not extract any work here, work could in principle be extracted from this photon-number difference for example by charging a capacitor as proposed in~\cite{Barbieri16MD}, or by moving a particle via the differential in radiation pressure thus engineered.
We consider $\nicefrac{\Delta N}{N}$ --- the change in photon number per incident photon --- as our figure of merit for the `power' of the MD.
Here, $N$ is the number of photons incident in mode $\mathrm{In_A}$ which we will always set equal to the number of photons in mode $\mathrm{In_B}$. We estimate $N$ by making the feed-forward inactive, measuring the photon number with the detectors in modes $D_A$ and $D_B$, and then dividing by the beamsplitter reflectivity.
We begin by presenting our derivations of this figure of merit for the different input states in the regime of low photon-numbers, which we define as one- and two-photon detection events.

\textit{Uncorrelated thermal states.} --- 
We first consider two uncorrelated thermal states with the same photon number, as in Fig.~\ref{fig:md_mosaic} \textbf{(e)}.
%= \left(1-e^{-\beta\hbar\omega}\right)\sum_{n=0}^{\infty}e^{-n\beta\hbar\omega}\ket{n}\bra{n}
The density matrix of a single-mode thermal state of temperature $T$ ($\beta = \nicefrac{1}{(k_{b}T)}$) is $\rho_\mathrm{Th} =\sum_{n=0}^\infty P_n\ket{n}\bra{n},$
where $P(n)=\overline{n}^n\left((1+\overline{n})^{n+1}\right)^{-1}$ and $\overline{n}=(e^{\beta\hbar\omega}-1)^{-1}$ is the average photon number in this state ($\hbar$ and $k_B$ are the Planck and Boltzmann constant, respectively and $\omega$ is the angular frequency of the photon).
This state is sent into mode $\mathrm{In_A}$ towards the beamsplitter, with vacuum incident into the other unlabelled mode.
At the beamsplitter, some of the light is reflected towards detector $\mathrm{Dem}_A$ while the rest is transmitted into mode $A$.
The probability to detect $n$ photons at detector $\mathrm{Dem}_A$ and $m$ photons in mode $A$ is  $P(m,n)=\frac{(n+m)!}{n!m!}\frac{\overline{n}^{n+m}}{(1+\overline{n})^{n+m+1}}R^{2n}(1-R^2)^{m}$,
where $R$ is the reflection amplitude of the beamsplitter \cite{loudon2000quantum}.
The subtraction of one photon from a thermal state leads to a two-fold increase in the average photon number left in mode $A$,  $\overline{n}'=2\overline{n}$ \cite{quantumtheory_continuous_photodetection} (hence the two red photons in Fig.~\ref{fig:md_mosaic} \textbf{(e)}). Moreover, if the initial thermal state is a single-mode thermal state, the remaining state is a multi-mode thermal state \cite{Bogdanov2017Multiphoton}.
The same is true for the lower $B$ modes.
Importantly, neglecting this detection event results in a thermal state with a lower average photon number $\overline{n}'=(1-|R|^2)\overline{n}$ in mode $A$ or $B$ \cite{loudon2000quantum}; hence, the MD's measurement does not inject any energy into either mode.
As a result, if the MD detects a photon at detector $\mathrm{Dem}_A$ the remaining photon number in mode $A$ is higher.
 Since the MD will attempt to increase the photon number in mode $D_A$, it will set the UFOS in the `bar state' so that light incident in mode $A$ ($B$) is transmitted to mode $D_A$ ($D_B$).
However, detection of a photon at detector $\mathrm{Dem}_B$ means that there are more photons in mode $B$, so the MD will swap modes $A$ and $B$ by setting the UFOS to the `cross state'.
When no photons are detected, or a photon is detected in each mode, the MD sets the UFOS to the `bar state'.
Thus  a photon number imbalance can be generated based on the MD's information (click/no-click), which depends on the beamsplitter's reflection amplitude $R$.
In Appendix~\ref{app:MDP}, we derive the power of the MD when using uncorrelated thermal states for low photon-numbers, obtaining
\begin{equation}\label{eq:DemonsPower_thermal}
    \frac{\Delta N}{N}=\frac{2\overline{n}}{(1-\overline{n})^2}R^2(1-R^2).
\end{equation}
Note that here $\overline{n}$ is the expectation value of the photon number per mode (i.e., the number of photons per coherence time, which is $\tau_c=5.42 \mu$s for our uncorrelated thermal photons and $\tau_c\approx2$ ps for the photons generated by spontaneous parametric down-conversion). In contrast, $N$ is the total number of photons sent into mode $\mathrm{In_A}$ (and equally into mode $\mathrm{In_B}$) during the experiment.
Hence, $\overline{n}=\frac{\tau_C}{T}N$, where $T$ is the total measurement time.
From Eq. \ref{eq:DemonsPower_thermal}, we see that the power of the MD increases with $\overline{n}$ (in the limit of $\overline{n}\ll 1$), indicating that this effect is driven by the thermal nature of the beams.
Since Eq.~\ref{eq:DemonsPower_thermal} also depends on the reflection amplitude $R$, we will tune this parameter to optimise the power of the MD. 

\textit{Split thermal states.} --- 
Now, if instead of two statistically uncorrelated thermal states we send two classically correlated thermal states (created by splitting a single thermal state at a balanced beamsplitter) into the setup, the state in modes $\mathrm{In_A}$ and $\mathrm{In_B}$ is $\rho_\mathrm{Split}=\sum_{n,m=0}^\infty P(n,m)\ket{n,m}\bra{n,m}_{\mathrm{In_A},\mathrm{In_B}}$, with $P(n,m)$ as defined above (for $R=\nicefrac{1}{\sqrt{2}}$).
Since the thermal states in each mode originated from the same source, a detection at $\mathrm{Dem}_A$ (or equivalently $\mathrm{Dem}_B$) affects the state in both modes $A$ and $B$: this amounts to effectively increasing the photon number in both arms simultaneously (Fig.~\ref{fig:md_mosaic} \textbf{(f)}), as in~\cite{QuantumVampireKatamadze:18}.

Therefore, when operating with split thermal states, the MD cannot possibly know in which arm the bunching effect will be stronger --- its power in this case is zero: $\nicefrac{\Delta N}{N}=0$, for all values of the reflectivity of the MD's beamsplitter (see Appendix~\ref{app:MDP}).
This result holds in general, not just in the regime of low photon-numbers~\cite{Theo_shu_MD}.
In terms of the partitioned box example, we can imagine a hole in the partition through which particles can freely escape to either side such that, when the MD measures one particle, the temperature is increased on both sides of the partition (Fig.~\ref{fig:md_mosaic} \textbf{(b)}). This is not a perfect analogy, as in our photonic implementation, particles are not exchanged after the MD's measurement. Rather, the thermal state is distributed among two optical modes and photon subtraction in either of these modes increases the average photon number equally in both modes, as in~\cite{QuantumVampireKatamadze:18} for example.

\textit{Correlated thermal states.} --- Unlike the above correlations, which effectively make the MD powerless, the correlations generated by type-II spontaneous parametric down conversion (SPDC) increase its power \cite{Theo_shu_MD}.
Ideally, for this state whenever there is a photon in mode $\mathrm{In_A}$, a twin photon is found in mode $\mathrm{In_B}$ (correlated particles are denoted as the red and blue particles in Fig. \ref{fig:md_mosaic} \textbf{(c)} and \textbf{(g)}).
In particular, the ideal state is $\ket{\psi}=\frac{1}{\cosh{s}}\sum_{n=0}^\infty(\tanh{s})^n\ket{n,n}_{\mathrm{In_A},\mathrm{In_B}}$, where $s$ specifies the brightness of the source.
Although photons are always emitted in pairs, tracing over photons in mode $\mathrm{In_A}$ ($\mathrm{In_B}$) leaves photons in mode $\mathrm{In_B}$ ($\mathrm{In_A}$) in a thermal state~\cite{loudon2000quantum}, which means that the power of the MD could be enhanced as with uncorrelated thermal states.
However, we will now see that the correlations generated by SPDC have a much larger effect, and in fact counteract the thermal bunching.

The key resource for this state are the correlations which provide the MD with more information than it can obtain from the thermal bunching.
For example, if exactly one photon is detected at $\mathrm{Dem}_A$, the MD has removed one photon from mode $A$, and it now knows that there will be one more photon in mode $B$ (and vice-versa).
This is illustrated Fig.~\ref{fig:md_mosaic} \textbf{(g)}, where the photons come in pairs. 
For every red photon in mode $\mathrm{Dem}_A$ there is a correlated blue photon in mode $B$. 
Hence, when the MD detects a photon in $\mathrm{Dem}_A$ there are now zero photons in mode $A$, and exactly one photon in mode $B$.
Therefore, the MD will move the remaining photon from $B$ to mode $D_A$ by putting the UFOS in the cross state. 
Similarly, it will set the UFOS to the bar state when it detects a photon in mode $B$. As before, the UFOS remain in bar for all other detection events

In the partitioned box analogy, Fig.~\ref{fig:md_mosaic} \textbf{(c)} shows the MD gaining information about particles on both sides of the partition, by measuring a particle on only one side.
Here the colour denotes correlated pairs of particles. For every red (blue) particle on the left there is a corresponding blue (red) particle on the right moving in a mirrored direction.
Thus, if the MD finds a particle impinging on the aperture from one side, it measures (and destroys) that particle leaving its twin to arrive alone from the other side.

Notice that now the MD's logic is reversed compared to the uncorrelated thermal states. Previously, finding a particle on one side implied that there were more particles on that side.
Now the opposite is true.
This also introduces a new experimental issue: the power of the MD critically depends on the ratio between correlated photons (which experimentally correspond to coincidence events) and uncorrelated photons (by which we mean photons that have lost their partner due to experimental loss).
In practice that means that the coupling efficiency $\epsilon^2$, defined as the ratio between the singles rate the coincidence rate, is essential to create a powerful demon.

In Appendix~\ref{app:MDP}, we derive the power of the MD using correlated thermal states in the regime of low photon numbers in the presence of experimental loss.
We obtain
\begin{equation}\label{eq:spdcSING_text}    \frac{\Delta N}{N}=2\epsilon^2R^2(1-R^2).
\end{equation}
This equation resembles Eq.~\eqref{eq:DemonsPower_thermal}.
Here, however, the power of the MD does not depend on $\overline{n}$, since the correlations are the driving force of the power of the demon in this regime.
In fact the correlations counteract the thermal bunching effect.
Moreover, notice that the power of the MD is scaled by the coupling efficiency $\epsilon^2$.
For a perfect source $\epsilon^2=1$.
Interestingly, we can experimentally probe this, by simply computing the power of the MD per photon pair (by normalising $\Delta N$ by the total number of coincidences $C$).
As we show in Appendix~\ref{app:MDP}, this results in:
\begin{equation}\label{eq:spdcCOIN_text}    \frac{\Delta N}{C}=2R^2(1-R^2).\end{equation}

\begin{figure*}[ht!]
    \centering\includegraphics[width=1\textwidth]{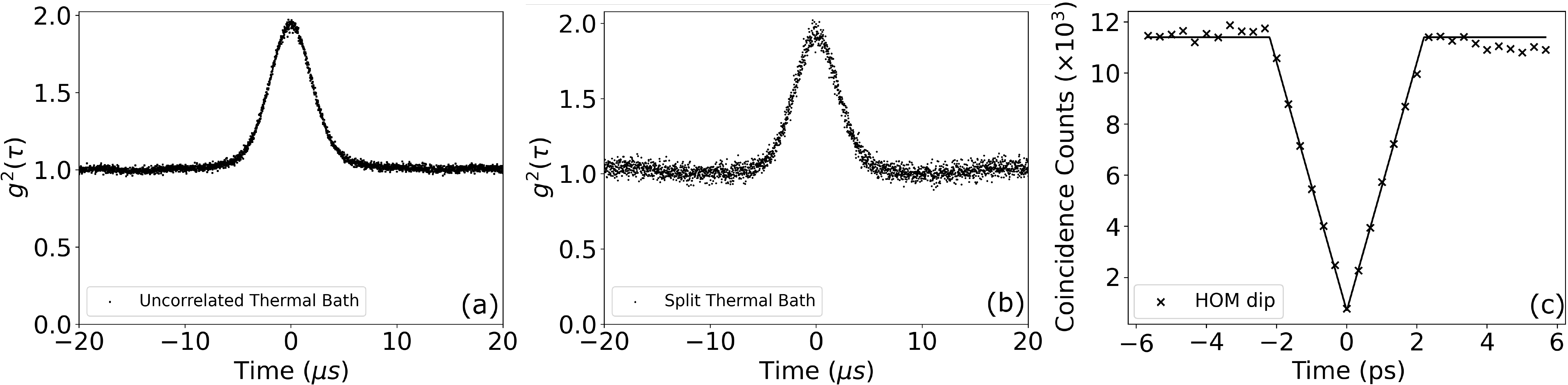}
    \caption{\textbf{Source characterisation.} Representative measured $g^{(2)}(\tau)$ of the thermal beam \textbf{(a)} and one mode of the split thermal beam \textbf{(b)}. Both display $g^{(2)}(\tau=0)$ very close to $2$.  Panel \textbf{(c)} shows the Hong-Ou-Mandel dip (visibility $v^2=0.87 \pm 0.02$) between the signal and idler down-converted photons with act as our resource for the anti-correlated states.}
    \label{fig:g2}
\end{figure*}

\textit{Anti-correlated thermal states.}
Finally, we consider the operation of the MD with anti-correlated N00N states (here we set $N=2$)~\cite{kokCreation}.
We refer to this state as anti-correlated because whenever there is a photon in mode $\mathrm{In_A}$, there are none in mode $\mathrm{In_B}$ (see Fig.~\ref{fig:md_mosaic}h).
In Ref. \cite{Theo_shu_MD}, a specific two-mode state was proposed with two important properties. 
First, for every N-photon event, the two modes are in a N00N state --- where there are $N$ photons mode $A$ and zero in mode $B$, and vice-versa.
Second, tracing out either mode, leaves the remaining mode in a thermal state.
Experimentally, we do not create this state, but we note that some experimentally-realised N00N state sources may come close to these conditions \cite{afek2010high,rozema2014scalable}.
Instead, here we perform Hong-Ou-Mandel (HOM) interference between the two photons from the SPDC source to create a 2-photon N00N state.  
The higher-order modes will not be in a N00N state; rather, they will be in so-called Holland-Burnett states \cite{holland1993interferometric}.
Nevertheless, as we show in Appendix~\ref{app:MDP} for low-photon numbers, the demon effect is driven by the two-photon component and in this regime our source behaves almost exactly as the state proposed in \cite{Theo_shu_MD}.
Ideally, our state has only the vacuum component and the two-photon terms $(\ket{2,0}_\mathrm{\mathrm{In_A},\mathrm{In_B}} + \ket{0,2}_\mathrm{\mathrm{In_A},\mathrm{In_B}})/\sqrt{2})$.
Starting from this state, in Appendix~\ref{app:MDP} we derive the power of the MD in the presence of loss (coupling efficiency $\epsilon^{2}$) and for imperfect Hong-Ou-Mandel visibility ($v^2$), finding
\begin{equation}\label{eq:noonSING_text}
\frac{\Delta N}{N}=2\epsilon^2(2v^2-1)R^2(1-R^2).
\end{equation}
This differs from operation with correlated states (Eq.~\eqref{eq:spdcSING_text}) by a factor depending on the visibility $2v^2-1$.
Here, the key resource are the anti-correlations: In Fig.~\ref{fig:md_mosaic} \textbf{(g)} photons come from the source in pairs, denoted by red and blue.
If the MD detects, say, exactly one photon at Dem$_A$, then it knows that a second photon is left in mode $A$, and there are no photons in mode $B$.
In analogy, Fig.~\ref{fig:md_mosaic} \textbf{(d)} shows the MD gaining information about both particles by measuring only one of them: if the MD detects, say, the red particle then it knows that another one is also arriving from the same side.
Here, as with the quantum correlated states above, the power of MD depends on the coupling efficiency, but also on the visibility (i.e. the quality of the N00N state source).
Again, if we instead normalise our results by the average coincidence rate $C$, we get closer to the ideal performance
\begin{equation}\label{eq:noonCOIN_text} \frac{\Delta N}{C}=2(2v^2-1)R^2(1-R^2),\end{equation}
but this does not remove the visibility dependence.

Finally, we remark that in the regime of low photon numbers and for optimal visibility $v^2=1$, operation with correlated and anti-correlated states would provide the same power of the MD.
However, for anti-correlated states, the demon switches particles in the same manner as for the thermal bunching. This is why at higher temperatures (photon numbers) Ref.~\cite{Theo_shu_MD} found that the anti-correlated states greatly out perform the correlated states.

\section{Experiment}

Our photonic MD experiment, shown in Fig.~\ref{fig:setup}, is built around a 2x2 UFOS (BATi 2x2 Nanona fibre switch).
This optical switch can route light from two input modes into two output modes with a variable splitting ratio.
The response time of the UFOS is below 60 ns, with a maximal duty-cycle of 1 MHz, and a cross-channel isolation greater than 20 dB for any polarisation,
see Ref.~\cite{Zanin_fibre_compatible} for more details.
The reflectivity $R$ of the variable beamspliter, implemented by a TDC, sets the amount of information gained by the MD about the input state.
After the BS, photons go through a $152$ m fibre delay line \footnote{The extra fibre relaxes the constraints on the adjustment of electronic delays, pulse size, and other parameters that are more easily controlled than the physical length of optical fibres and BNC cable.} before entering the UFOS, where the path they take (towards either detector $D_A$ or $D_B$) is conditionally controlled by the MD.
All detectors are superconducting nanowire single-photon detectors (SNSPD) from PhotonSpot Inc. (deadtime 50 ns, average system detection efficiency 90\%, and timing jitter of $\approx 150$ ps).
Detection events are recorded and processed by a commercial time tagging module (UQDevices
Logic16 TTM).
A photon detection event triggers a wave-function generator (Keysight 33500B Series) to generate a $2.5~ \mu$s TTL pulse that will make the UFOS switch states.

Before presenting our experimental results, we will explain our measurement procedure which is used for the four different initial states.
Our target metric is $\Delta N$, which we then normalise by either the total singles count $N$ or the coincidence count $C$.  
However, this is highly sensitive to variations in the imbalance between photon numbers at the input, and thus also to losses in the setup, which can (and do) differ slightly between the upper $A$ modes and $B$ lower modes.
In order to account for these unbalanced losses, we first balance the count rates of the two arms when the UFOS is fixed in the bar state and then measure the counts at the output. This ensures that $\Delta N_{\mathrm{bar}}=N_{A}-N_{B} = 0$, where $N_A$ ($N_B$) refers to the total number of photons detected in mode $D_A$ ($D_B$).
Second, we measure the intrinsic imbalance of the setup by fixing the UFOS to the cross state, and preventing the MD from controlling the path by feed-forward. In this configuration we measure $\Delta N_{\mathrm{cross}}=N_{A}-N_{B}$, which represents the maximum imbalance the MD could `accidentally' achieve without properly using information from either thermal bunching or the correlations between the states.
We then measure the photon number difference with the MD being active, i.e. we turn on the feed-forward logic (which differs depending on the input state), yielding $\Delta N_\mathrm{FF}=N_{A}-N_{B}$. Given these measurements, we construct the unnormalised power of the MD as
\begin{equation}\label{eq:DeltaN_Demon}    \Delta N:= \Delta N_\mathrm{FF}-\Delta N_{\mathrm{cross}}. \end{equation}

Finally, we need to normalise this by either the number of input single photons N or the number of incident photon pairs C.
To estimate the N, we simply measure the total number of counts at the end of the experiment at the detectors in modes $D_A$ and $D_B$ for each reflectivity of the TDC.  We then divide these rates by the reflectivity to obtain the singles rates arriving at the beamsplitter, and average over the counts in modes $A$ and $B$.
To estimate the incoming coincidence counts, we perform a similar protocol. However, due to limitations in the counting logic, we only measured the coincidence rates between detectors $D_A$ and $\mathrm{Dem}_A$ ($C_{D_A,\mathrm{Dem}_A}$) and detectors $D_A$ and $\mathrm{Dem}_B$ ($C_{D_A,\mathrm{Dem}_B}$) for a reflectivity of $50\%$.  From these measurements we compute the input coincidence rates for both the correlated and anti-correlated input states (note that this calculation differs slightly for the N00N state and SPDC state).

\textit{Uncorrelated thermal state.} --- To generate uncorrelated thermal baths we make two independent thermal states of light, using a rotating ground glass wheel (see Appendix~\ref{app:exp}). 
We verify the thermal nature of the source by measuring $g^{(2)}(\tau=0)$, which is typically $\approx1.95$.
A representative measurement is presented in Fig. \ref{fig:g2} \textbf{(a)}.
Since $g^{(2)}(\tau=0)>1$ is only a necessary condition for a thermal state, we further verify that without the rotating ground glass $g^{(2)}(\tau=0)\approx 1$. With these measurements, and knowledge of our experiment, we can be confident that we have generated a thermal state.
By fitting to the $g^{(2)}(\tau)$, assuming a Gaussian thermal beam of the form $g^{(2)}(\tau)=1+e^{-\pi(\nicefrac{\tau}{\tau_c})^2}$~\cite{loudon2000quantum},
we obtain a coherence time of $\tau_c=5.42~\mu$s. 

\begin{figure*}[ht!]
    \centering\includegraphics[width=1\textwidth]{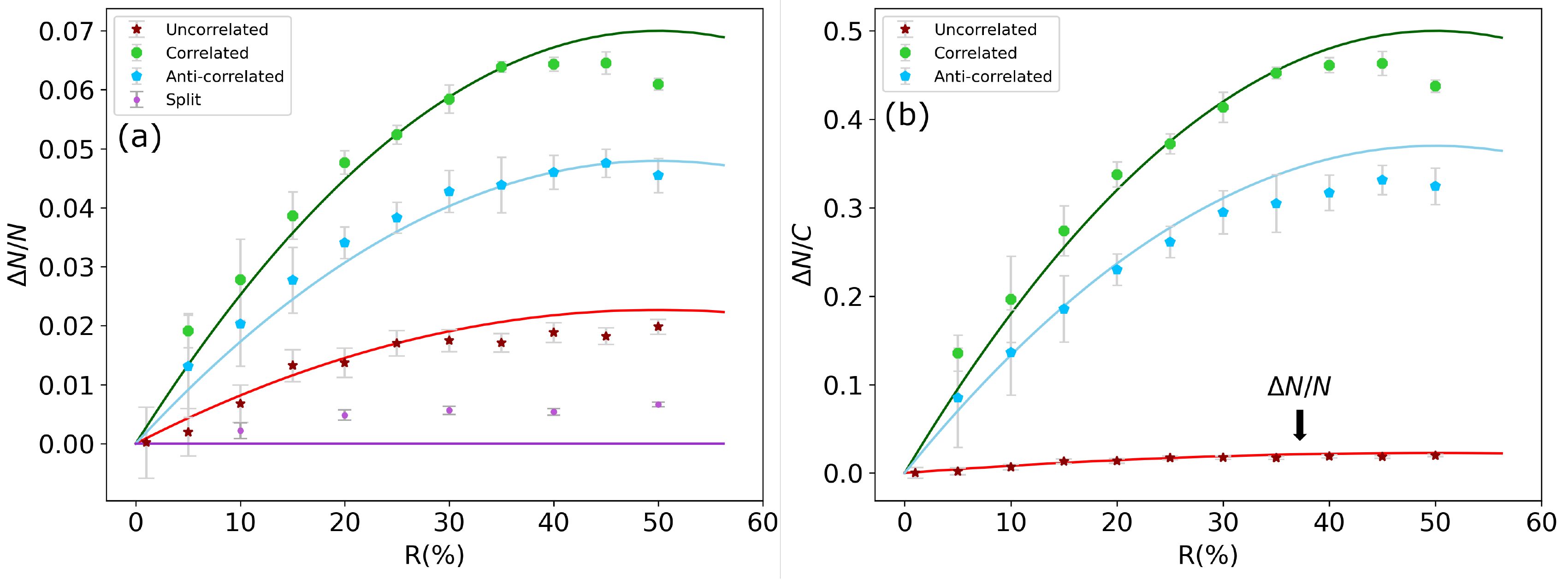}
    \caption{\textbf{Power of Maxwell's Demon.} \textbf{(a)}, The photon number difference, $\Delta N$ that the Demon can achieve per incident photon: Eq.~\eqref{eq:DeltaN_Demon} for various correlations. Solid curves are the theory for: uncorrelated state~\eqref{eq:DemonsPower_thermal} with $\overline{n}=0.05$ (red); split thermal state $\nicefrac{\Delta N}{N}=0$ (violet); correlated state~\eqref{eq:spdcSING_text} with $\epsilon^2=0.14$ (turquoise); anti-correlated state~\eqref{eq:noonSING_text} with $\epsilon^2=0.14$ and $v^2=0.87$ (green).
    \textbf{(b)}, The same  $\Delta N$ data, but here the correlated and anti-correlated data are normalised by the correlation rate to compute the demon's power per incident pair. {Note, the uncorrelated data plotted in red in this panel is normalised by the singles counts, and is included here for comparison}. The solid curves are the theory for: uncorrelated state~\eqref{eq:DemonsPower_thermal} with $\overline{n}=0.05$ normalised with coincidence rate (red); correlated state~\eqref{eq:spdcCOIN_text} (turquoise); anti-correlated state~\eqref{eq:noonCOIN_text} with $v^2=0.87$ (green).
    The experimental errors are the standard deviation of the measurement set.
    }
    \label{fig:md_cexpt}
\end{figure*}

We then proceed as described above, setting $\Delta N_{\mathrm{bar}}=0$  and measuring $\Delta N_{\mathrm{cross}}$ and $\Delta N_{FF}$ as we tune the MD's reflectivity from $0\%$ to $50\%$.
For every reflectivity we reconstruct the power of the MD, defined above.
The result is plotted in Fig.~\ref{fig:md_cexpt} as red stars.
For these measurements, we set the average incident photon number in modes $\mathrm{In_A}$ and $\mathrm{In_B}$ to $27,800 \pm 500$  cps.
We fit the experimental data with a single parameter, the average photon number per mode $\overline{n}$.
Given the $5.42\mu$s coherence time of the pseudo thermal light and the photon rate, we expect $\overline{n}=0.151\pm0.002$ (the error bars are the standard deviation of the measured singles rates for different reflectivities).
However, we find a much better fit with a value of $\overline{n}=0.05$.

The discrepancy in $\overline{n}$ is likely due to experimental limitations in our switching logic.
We set the switching time to $2.5\mu$s, which is less than the photon coherence time.
However, setting it higher limits our maximum count rate, since the wave-function generator cannot respond between time it is triggered and the output pulse is being generated.
Because of our reduced switching time, some bunched events may be missed, thus limiting the power of the MD.
We use this value in our theoretical calculation shown in Fig.~\ref{fig:md_cexpt} \textbf{(a)} (solid red curve).
Note, we also plot this data in Fig.~\ref{fig:md_cexpt} \textbf{(b)}.
Although it is experimentally possible for us to increase the average photon number, the long coherence time of our thermal distribution would then take us out of the regime of low photon numbers, which is the basis of the fair comparison of the power of the MD with other input states.

\textit{Split thermal state.} --- We produce this input state by simply splitting one of the thermal states at a balanced beamsplitter after the ground glass wheel.
We first verify that this still results in a thermal state in each mode by measuring the $g^{(2)}(\tau)$ after the beamsplitter, a representative $g^{(2)}(\tau)$ is plotted in Fig.~\ref{fig:g2} \textbf{(b)}, displaying a $g^{(2)}(\tau=0)$ close to $2$.

After this, we balance the count rates and measure the power of the MD as before.
The result is plotted in Fig.~\ref{fig:md_cexpt} \textbf{(a)} as the violet diamonds.
For these data, $\overline{n}\approx 0.7$ in each spatial mode at the input.
Although for this $\overline{n}$, our low-photon approximation no longer applies, Ref. \cite{Theo_shu_MD} showed that our result is independent of $\overline{n}$.
Experimentally, this corresponds to incident singles rates in modes $\mathrm{In_A}$ and $\mathrm{In_B}$ of $293,000\pm 3000$ cps.
The resulting experimentally measured $\nicefrac{\Delta N}{N}$ is very close to zero, as expected.
The small deviation from zero is likely caused by slightly imbalanced losses in the setup, such that although $\Delta N_{OFF}$ is close to $0$ the real photon number before the UFOS is not exactly balanced.

\textit{Correlated thermal states.} --- We produce correlated thermal states by SPDC.
In this source (described in Appendix~\ref{app:exp}) the coherence time is $\approx 2$ ps (which we verify by a Hong-Ou-Mandel experiment in Fig.~\ref{fig:g2} \textbf{(c)}).
This is much smaller than both the minimum resolution of our time tagger (which is $156$ ps), and the timing jitter of our detectors ($\approx 200$ ps).
Hence, we cannot measure the $g^{(2)} (\tau = 0)$ as the width of the expected peak is below the minimum resolution of our detection system.
However, it is now well established that in such sources the signal (idler) beam is indeed thermal when the idler (signal) beam is traced out~\cite{YurkeObtainment,BlauensteinerPhoton}.

For this state, we operate with a balanced average photon rate of $168,000\pm1000$ cps in modes $\mathrm{In_A}$ and $\mathrm{In_B}$.
Although this is a  higher count rate than for the uncorrelated states, the shorter $2$ ps coherence time means the average photon number per mode is much smaller $\overline{n}\approx 10^{-7}$.
We measure the power of the MD per incident photon by varying $R$ (green circles in Fig.~\ref{fig:md_cexpt} \textbf{(a)}).
In this case, the theoretical curve (Eq.~\eqref{eq:spdcSING_text}) depends only on the coupling efficiency $\epsilon^2$.
This is calculated by measuring the coincidence rate between the detectors and correcting for the reflectivity of each BS loss and dividing by the singles rate, yielding  $\epsilon^2=0.14$.
The resulting solid green curve fits the data very well.

In Fig.~\ref{fig:md_cexpt} \textbf{(b)}, we also plot the power of the MD per incident photon pair, by normalising  $\Delta N$ by the coincidence rate (green circles).
This removes the effect of almost all experimental imperfections, and our experimentally measured power of the MD almost reaches the theoretical maximum value of $0.5$ (see Eq.~\eqref{eq:spdcCOIN_text}).
Moreover, the experimentally measured data is in good agreement with the full theoretical curve (solid green line).

\begin{figure*}[ht!]
    \centering\includegraphics[width=1\textwidth]{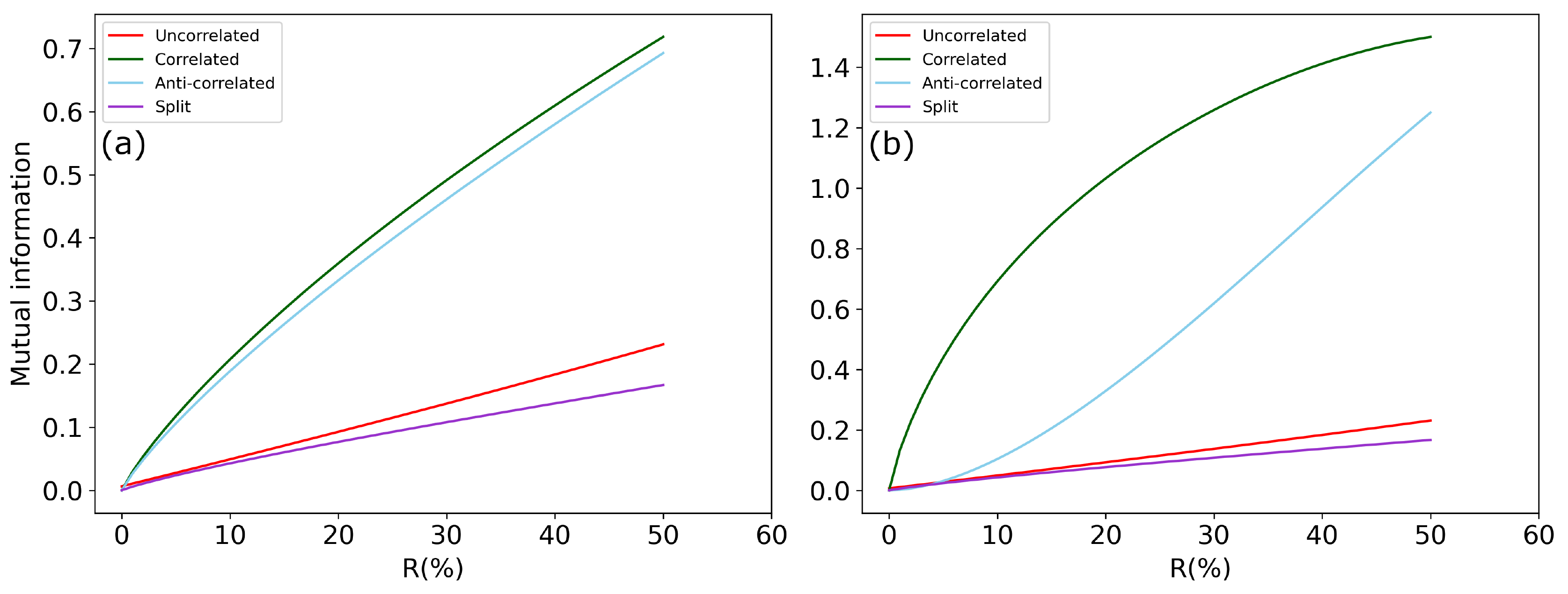}
    \caption{\textbf{MD's Information.} The mutual information between the MD's measurement outcomes and the photon number a) for our experimental parameters, and b) for the ideal situation.
    The parameters used to evaluate these curves are the same as those used in the fits to the data presented in Fig. \ref{fig:md_cexpt}.
    In all cases, as the MD's beamsplitters reflectively increases, it gains more information about the photon number.
    Moreover, for all of the parameter regimes investigated in this work the correlated and anti-correlated states out perform the uncorrelated thermal states.
    }
    \label{fig:md_info}
\end{figure*}

\textit{Anti-correlated thermal states.} --- We generate the input 2-photon N00N state by HOM interference between the two photons of our SPDC source.
The details are described in Appendix~\ref{app:exp}. 
We estimate the quality of the N00N state by scanning the relative delay and measuring the visibility of the HOM dip --- we measure $v^2=0.87 \pm 0.02$.
To ensure that our state is classically anti-correlated in the Fock basis, we observe the absence of correlations (coincidences) at a time delay of zero (Fig.~\ref{fig:g2} \textbf{(c)}), which implies that the two photons incident on the fiber beamsplitter in Fig.~\ref{fig:setup} \textbf{(c)} bunch and exit in the same mode.
As mentioned above, this does not result in thermal states in modes $A$ and $B$, but we show in Appendix~\ref{app:MDP} that the behaviour is almost identical in the low-photon number regime.

For this state, we operate with balanced singles rates of $141,000\pm3000$ cps, and, again, $\overline{n}\approx10^{-7}$.
Proceeding as before, we obtain the power of the MD and plot the result normalised by singles  as turquoise pentagons in Fig.~\ref{fig:md_cexpt} \textbf{(a)}.
For the corresponding theory curve (Eq.~\eqref{eq:noonSING_text}), we use our measured HOM visibility of $v^2=0.87$ and take $\epsilon^2=0.14$ (as above).
In Fig.~\ref{fig:md_cexpt} \textbf{(b)}, we normalise $\Delta N$ by the measured coincidence rate.
The theory curve, Eq.~\eqref{eq:noonCOIN_text}, only requires the visibility which we take as above, and plot as the solid turquoise line.
Although ideally the correlated and anti-correlated states behave identically for low photon numbers, the anti-correlated states suffer an additional imperfection arising from the HOM interference.

\section{Discussion}
Firstly, we note that the agreement between the experimental data and the theory is very good for all states, with a slight deviation at high reflectivity.
This deviation comes from our simplified theory, which does not include the MD's detection efficiency.
Indeed, in the theory of~\cite{Theo_shu_MD} it can be observed that changes in the relative detection efficiency between detectors $\mathrm{Dem}_A$ and $\mathrm{Dem}_B$ can shift the dependence of $\Delta N$ on reflectivity.
Moreover, we stress that only one data set, that of the uncorrelated thermal states normalised by singles, required a fitting parameter.
All other curves presented in Fig.~\ref{fig:md_cexpt} agree with experimental data using only independently measured experimental parameters, such as the coupling efficiency and HOM visibility.
This demonstrates the great potential of relatively simple optical experiments to investigate information processing scenarios characteristic of thermodynamics using pseudo-thermal beams, which moreover give access to a wealth of exotic correlations readily obtained with photons.

We have experimentally demonstrated that the power of the MD with correlated and anti-correlated baths (which we produce with entangled quantum states) vastly outperforms operation with classical states (uncorrelated and split thermal states),  as predicted in~\cite{Theo_shu_MD}.
For example, when considering the power of the MD per photon (i.e. the data normalised to the incident single-photon rate, plotted in Fig.~\ref{fig:md_cexpt} \textbf{(a)}), the power of the MD is a factor of $\approx 5$ higher with correlated baths than with uncorrelated baths {(green curve compared to the red curve in panel \textbf{(a)}}.
Moreover, the enhancement is even more striking when considering the power of the MD per photon pair, plotted in Fig.~\ref{fig:md_cexpt} \textbf{(b)}.
This analysis serves to essentially filter out the single photons that have lost their partner, and allows us to reach the theoretical limits for the correlated baths.
In particular, although operating with uncorrelated thermal states does yield a finite power, this is outperformed by a factor 16 when operating with anti-correlations {(blue curve compared to the red curve in panel \textbf{(b)})}, and by a factor 28 when operating with correlated states {(green curve compared to the red curve in panel \textbf{(b)})}.
Here we are comparing to the power of the MD with uncorrelated thermal baths normalised by the singles rate {(plotted in red in both panels \textbf{(a)} and \textbf{(b)})}, since the thermal state is robust to loss.

The MD's ability to extract work is limited by the amount of information it can extract.
While this was originally formalized by Landauer's erasure principle \cite{Landauer_irrev_1961, bennettThermodynamicsComputationReview1982}, applying this concept in a quantitative manner to our system would require a generalized Jarzynski equality such as that derived in \cite{sagawaGeneralizedJarzynskiEquality2010}.
Nonetheless, the mutual information  $I$ between the MD's measurement outcomes in modes $\mathrm{Dem}_A$ and $\mathrm{Dem}_B$ and the photon number in mode $A$ and mode $B$ can be used to qualitatively explain the enhancement of the power of the MD in the presence of the correlated input states.
In Appendix~\ref{app:info}, we derive $I$ for each input state, and we plot the results in Fig.~\ref{fig:md_info}.
Note that MD has 4 measurement outcomes so the maximum possible entropy is 2. 
In panel \textbf{(a)} we have used the coupling efficiency and N00N state visibility used to fit to our experimental data normalized by the singles, while in panel \textbf{(b)} we used the parameters used to fit to the data normalized by the coincidence rate. In both panels $\overline{n}=0.05$ for the uncorrelated and split thermal cases.
These plots demonstrate that, for the both the correlated and anti-correlated states, the MD has access to more information, which is consistent with our observation that it can generate a larger $\Delta N$ for these data.
Interestingly, for the split thermal state the MD only has marginally less information than it does for the uncorrelated states.
It is hence possible that, for an operation that is more complex than simply the swapping the modes, the MD could still operate with this state.

In our analysis we have not considered the absolute efficiency of the MD, but rather focused on the enhancement provided by correlations in the baths.
In the original thought experiment, the cost of both the demon’s measurement of the particle’s velocity and its cost to open the aperture was assumed to be negligible.
This was later formalized by Landauer, who showed that if a reversible process is used for the measurement and sorting, then the cost (i.e. entropy increase) from this step is zero.
He then famously showed that at some point the MD must erase its measurement results, which will increase the entropy \cite{Landauer_irrev_1961, bennettThermodynamicsComputationReview1982}.
In our experiment, we used irreversible measurements, and hence we are not in the idealized regime. 
In principle, however, a photonic platform could reach this regime.
For example, the photon that the demon subtracts could be used to trigger a unitary (and hence reversible) controlled-path gate \cite{zhou2011adding} that operates on multi-particle input states.
Nevertheless, the measurement and switching in our experiment do not inject any external energy into the baths.
Note that the MD's beamsplitters do extract energy from the baths, but since this extraction is balanced in Mode $A$ and Mode $B$ this is not the source of the power of the MD.
Thus the enhancement we observed here would also be present in the ideal scenario.

Our work is valid for low photon numbers (\textit{i.e.} low temperature of the thermal beams).
For higher photon numbers, we expect the anti-correlated baths to perform the best, since the anti-correlations act in the same `direction' as the thermal correlations.
Thus a natural extension of our work would be to create a superposition of N00N states for various N~\cite{Theo_shu_MD}. To fully take advantage of this effect, one would need to create such states with a higher $\overline{n}$, which would likely require the use of a pulsed laser when generating the photon pairs by SPDC.

To conclude, we have implemented the first photonic Maxwell's Demon without post-processing, and verified that operating with correlations enhances the power of the MD to process the information it gains and generate a temperature difference (which enables the extraction of work) through classical feed-forward by over an order of magnitude.
Our work lays the foundations for future investigations of the interrelation between quantum information processing and thermodynamics at the microscopic scale.
For example our work calls for a theoretical derivation of the fluctuation relations~\cite{sagawaGeneralizedJarzynskiEquality2010} that could describe the `experimental information' obtained by the MD in each round of our feed-forward experiment.
Finally, although we used entangled states to generate correlations, our MD protocol only required classical correlations.  It would be interesting to develop MD protocols that exploit the quantum entanglement that is readily available in our system.

\section*{Data Availability}
All of the data that are necessary to replicate, verify, falsify and/or reuse this research is available online at~\cite{zanindata}.

\section*{Acknowledgements}
We are thankful to Beate Asenbeck and Giulia Rubino for their valuable comments on the manuscript.
We also thank Michele Spagnolo for work on the SPDC source, and Irati Alonso Calafell for help with the operation of the SNSPDs. The authors acknowledge the excellent comments given by the reviewers.
This work was supported by Austrian Science Fund (FWF): F 7113-N38 (BeyondC), FG 5-N (Research Group), P 30817-N36 (GRIPS) and P 30067\_N36 (NaMuG);
{\"O}sterreichische Forschungsf{\"o}rderungsgesellschaft (QuantERA ERA-NET Cofund project HiPhoP (No.731473)); 
Research Platform for Testing the Quantum and Gravity Interface (TURIS), the European Commission (ErBeSta (No.800942)), Christian Doppler Forschungsgesellschaft; {\"O}sterreichische Nationalstiftung f{\"u}r Forschung, Technologie und Entwicklung; Bundesministerium f{\"u}r Digitalisierung und Wirtschaftsstandort. GLZ and PHSR acknowledge funding from the Conselho Nacional de Desenvolvimento Científico 
e Tecnológico, CNPq (204937/2018-3 and 304123/2017-0), Coordena\c c\~{a}o de Aperfei\c coamento de Pessoal de N\'\i vel Superior (CAPES) and National Institute of Science and Technology for
Quantum Information, INCT-IQ.

\bibliographystyle{unsrtnat}%{quantum}%{unsrt}quantum
\bibliography{bibliography.bib}

\clearpage
\onecolumn
\appendix
\section{Experimental setup}\label{app:exp}
In this section, we detail the arrangement of the experimental setup and provide all technical information needed to reproduce it.

\subsection{Source of uncorrelated thermal light}
We produce two single-mode pseudo-thermal states with the Arecchi’s wheel~\cite{Martienssen1964_coherenceFluctuations,Arecchi1965,ARECCHI1966} setup shown in Fig.~\ref{fig:setup} \textbf{(a)}.
A laser beam of $\lambda=$1550 nm is divided into two spatial modes by a fibre beamsplitter (FBS).
These modes are focused (lens L1, f=75mm) on two separate spots on the spinning ground glass (grit size 220).
The rotation frequency of the step-motor (SM) is 20Hz.
A bare fibre after the ground glass collects thermal light in each mode.
We measure the second-order correlation $g^{(2)}{(\tau)}$ by the Brown and Twiss method~\cite{BROWN1956,TWISS1957}, and obtain $g^{(2)}(\tau)\approx 1.95$ (to be compared with $g^{(2)}(\tau) = 2$ for thermal light) and a  coherence time $\tau\approx 5.42$ $\mu s$.

\subsection{Source of correlated thermal states}
A 775 nm, CW beam is focused (lens L2, f=400mm) on a 30mm long ppKTP crystal in a Sagnac-loop configuration \cite{Jin:14PPkptSagnac}, yielding the singlet state $\ket\Psi^-$.
In Fig.~\ref{fig:setup} \textbf{(b)}, a dichroic mirror (DM) reflects 775 nm light and transmits 1550 nm.
We use standard mirrors (M), quarter- and half-waveplates (QWP, HWP) and polarising beamsplitters (PBS).
The output light is sent to fibre couplers (FC).

\subsection{Generation of anti-correlated thermal states}
The Hong-Ou-Mandel (HOM) interferometer presented in Fig.~\ref{fig:setup} \textbf{(c)} takes photons from the SPDC source (Fig.~\ref{fig:setup} \textbf{(b)}) and produces anti-correlated photons by spatially and temporally overlapping the modes with a delay line (DL) and a FBS, respectively.

\subsection{Operation of the photonic MD}

\begin{figure*}[ht]
    \centering\includegraphics[width=\textwidth]{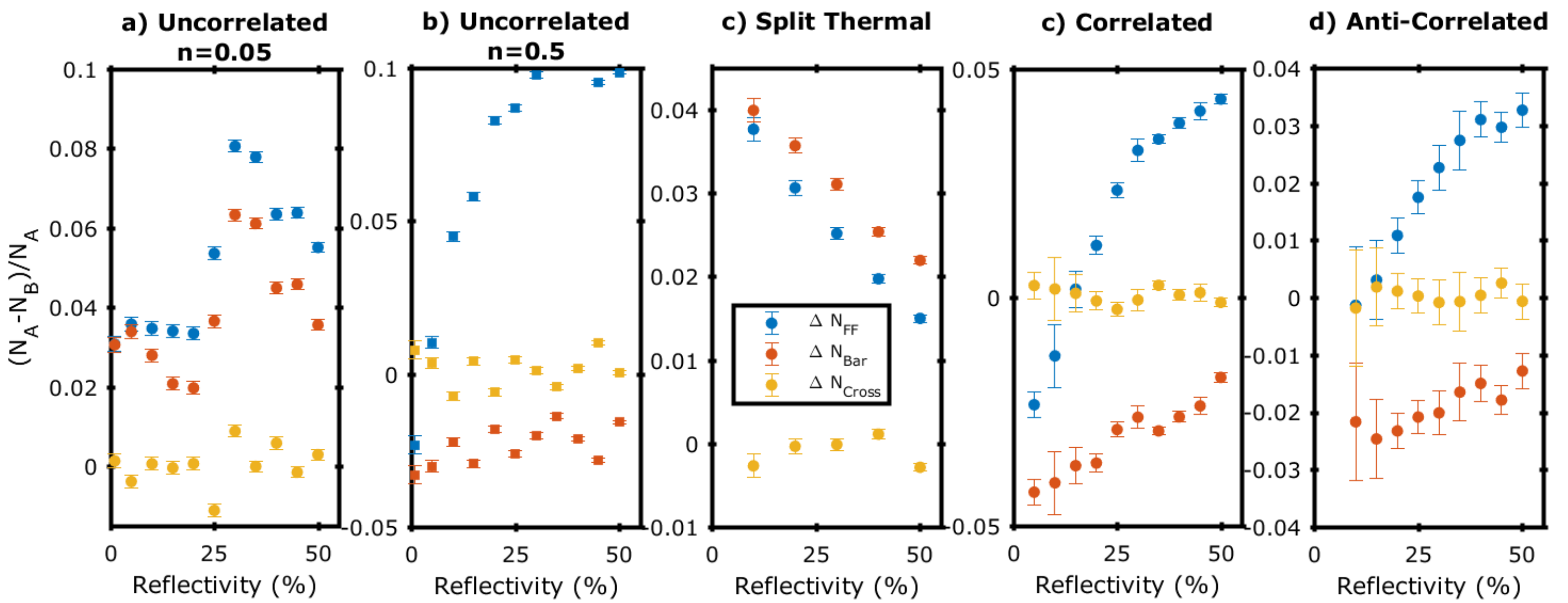}
    \caption{Individual Terms to compute the power of the MD. Here we plot the measured photon differences normalized by the input photon number when the UFOS is set to bar $\Delta N_\mathrm{bar}$ (orange), cross $\Delta N_\mathrm{cross}$ (red), and when the feed forward is active $\Delta N_\mathrm{FF}$ (blue). Data are shown for the uncorrelated thermal state with $\overline{n}=0.05$ (Panel a) and $\overline{n}=0.5$ (Panel b), split thermal state (Panel c), correlated state (Panel d), and the anti-correlated state (Panel e). }
    \label{fig:delta}
\end{figure*}

The active MD is shown in Fig.~\ref{fig:setup} \textbf{(d)}.
The input statistics are fed in the setup simply by connecting the fibres from either of the sources (Fig.~\ref{fig:setup} \textbf{(a)}, \textbf{(a)}+BS, \textbf{(b)}, \textbf{(b)+(c)}).
At the TDC's, the reflected arm goes to the Demon detectors (Dem$_{A,B}$), the transmitted arm goes to a fibre delay (FD) and then goes to the UFOS, whose configuration (`cross' or `bar' state) is decided conditionally on the MD's information.
We can adapt the switching logic according to the statistics of the MD's states.
After the UFOS, photons are sent to detectors $D_{A,B}$ and the output statistics are recorded with the time tagging module (TTM)
The photons' polarisation state is maintained with fibre paddles (PD) before and after the UFOS.
 
\section{Switching logic for the feed-forward}\label{app:ufos}
Here we detail the switching logic used in the feed-forward control of the UFOS state.
When this logic is applied, the difference (power of the MD) $\Delta N=N_{A}-N_{B}$ is improved by using the MD's information(Dem$_{(A,B)}$).
\begin{table}[ht]
\centering
\begin{tabular}{|c|c|}
\hline
\textbf{Dem(A,B)} & \textbf{UFOS state} \\ \hline
Dem(0,0)          & bar                                  \\ \hline
Dem(1,1)          & bar                                  \\ \hline
Dem(1,0)          & bar                     \\ \hline
Dem(0,1)          & cross                    \\ \hline
\end{tabular}
\caption{\textbf{Logic for uncorrelated thermal states, split thermal states and anti-correlated states.} Dem(A,B) the MD's detectors -- 0 (1) represents \textit{no click} (\textit{click}). Bar state -- input 1 (2) goes to output 1 (2); cross state -- input 1 (2) goes to output 2 (1).}
\label{tab:logic_demon}
\end{table}

\begin{table}[ht]
\centering
\begin{tabular}{|c|c|}
\hline
\textbf{Dem(A,B)} & \textbf{UFOS state} \\ \hline
Dem(0,0)          & bar                                 \\ \hline
Dem(1,1)          & bar                                  \\ \hline
Dem(1,0)          & cross                                  \\ \hline
Dem(0,1)          & bar                                  \\ \hline
\end{tabular}
\caption{\textbf{Logic for correlated thermal states.} Details as in Tab.~\ref{tab:logic_demon}.}
\label{tab:spdc_logic}
\end{table}

\section{Derivation of the power of the MD}\label{app:MDP}
Here we will present our simple calculations for our photonic Maxwell's Demon in the low-photon number regime.
In all cases, we define this low-photon regime to be events with two or less photons.
Note that although some intermediate equations in the following derivations contain contributions from higher-order terms, all of our final results depend only on one- and two-photon detection events.

\subsection{Uncorrelated Thermal States}
A thermal state of light is described by:
\begin{equation}\label{eq:rhothermalap}
    \rho_\mathrm{Th}=\sum_{n=0}^\infty P_n\ket{n}\bra{n},
\end{equation}
where 
\begin{equation}\label{eq:pnap}
    P(n)=\frac{\overline{n}^n}{(1+\overline{n})^{n+1}}.
\end{equation}
here $\overline{n}$ is the average photon number per mode.
Sending this state to a beamsplitter (with vacuum incident in the second input port) that reflects the light into mode $\mathrm{Dem}_A$, and transmits it into mode $A$ (as sketched in Fig. \ref{fig:md_mosaic}e,  where mode $A$ refers to the mode before and after the beamsplitter), results in the following probabilities to detect n photons in mode $\mathrm{Dem}_A$ and m photons in mode $A$ \cite{loudon2000quantum}:
\begin{equation}\label{eq:pnmap}
    P(m,n)=\frac{(n+m)!}{n!m!}\frac{\overline{n}^{n+m}}{(1+\overline{n})^{n+m+1}}R^{2n}(1-R^2)^{m},
\end{equation}
where $R$ is the reflection \emph{amplitude} of the beamsplitter.
Then after the beamsplitter, we will approximate the state to be:
\begin{eqnarray}
    \sigma_{\mathrm{Dem}_A,A}&=&P(0,0)\ket{0,0}\bra{0,0}_{\mathrm{Dem}_A,A} +
    P(1,0)\ket{0,1}\bra{0,1}_{\mathrm{Dem}_A,A} +\\\nonumber  &&P(0,1)\ket{1,0}\bra{1,0}_{\mathrm{Dem}_A,A} + P(1,1)\ket{1,1}\bra{1,1}_{\mathrm{Dem}_A,A} +\dots
\end{eqnarray}
In this expression we have ignored all off-diagonal terms, and terms with more than two photons.
Furthermore, we have left out the terms $P(2,0)\ket{0,2}\bra{0,2}_{\mathrm{Dem}_A,A}$ and $P(0,2)\ket{2,0}\bra{2,0}_{\mathrm{Dem}_A,A}$ since $P(2,0)<<P(1,0)$ for low photon numbers. This condition is valid when the probability of populating terms with photon numbers of three or higher is negligible compared to the one and two-photon contributions.
In practice this holds for $\overline{n} \lesssim 0.2$.
Then for our experiment we have one such state in modes $A$ and $\mathrm{Dem}_A$, and one in modes $B$ and $\mathrm{Dem}_B$. Thus the total input state is $\sigma_{\mathrm{Dem}_A,A}\otimes\sigma_{\mathrm{Dem}_B,B}$, with $\sigma_{i,j}$ defined above.
The overall expression is straight-forward but rather lengthy, so we do not include it here.

The next step is to include the action of the demon.
For uncorrelated thermal states, when the demon detects a photon in mode $\mathrm{Dem}_A$ ($\mathrm{Dem}_B$), the photon number in mode $A$, ($B$) is increased.
Since the demon's goal is to increase the temperature (photon number) of mode A, this means that when it detects a photon in $\mathrm{Dem}_A$ it does nothing. 
On the other hand, when it detects a photon in mode $\mathrm{Dem}_B$ there are more photons in mode $B$, so it should swap modes $A$ and $B$.
If the demon detects a photon in both mode $\mathrm{Dem}_A$ and mode $\mathrm{Dem}_B$ it gains no information; hence, it does nothing.
To represent this mathematically we simply make the following substitutions:
\begin{eqnarray}\label{eq:demonTherm}
    \ket{n,m,0,1}_{A,B,\mathrm{Dem}_A,\mathrm{Dem}_B} \rightarrow
   \ket{m,n,0,1}_{D_A,D_B,\mathrm{Dem}_A,\mathrm{Dem}_B}
\end{eqnarray}
Since we do not use number resolving detectors, we do not swap modes $A$ and $B$ in situations such as $\ket{n,m,1,2}$, as this will register as a detection event at both $\mathrm{Dem}_A$ and $\mathrm{Dem}_B$.
With this operation, the low-photon number state after the demon is approximated as

\begin{eqnarray}\label{eq:fullstate}\nonumber
\rho_\mathrm{out}= &
    P^2(0,0)\ket{0000}\bra{0000} + 
    P(0,0)P(0,1)\ket{0001}\bra{0001} +  P(0,0)P(1,0)\ket{0100}\bra{0100} + \\\nonumber &  P(0,0)P(1,1)\ket{1001}\bra{1001} +
    P(0,1)P(0,0)\ket{0010}\bra{0010} + 
    P^2(0,1)\ket{0011}\bra{0011} + \\\nonumber & P(0,1)P(1,0)\ket{0110}\bra{0110} +  P(0,1)P(1,1)\ket{0111}\bra{0111} + \\\nonumber &
    P(1,0)P(0,0)\ket{1000}\bra{1000} + P(1,0)P(0,1)\ket{0101}\bra{0101} + 
    P^2(1,0)\ket{1100}\bra{1100} +  \\\nonumber &
    P(1,0)P(1,1)\ket{1101}\bra{1101} +
        P(1,1)P(0,0)\ket{1010}\bra{1010} + P(1,1)P(0,1)\ket{1011}\bra{1011} +\\\nonumber & P(1,1)P(1,0)\ket{1110}\bra{1110} +  P^2(1,1)\ket{1111}\bra{1111} + \dots\\
\end{eqnarray}

In Eq. \ref{eq:fullstate}, the mode labels are left off, but they run $D_A$, $D_B$, $\mathrm{Dem}_A$, $\mathrm{Dem}_B$; i.e. $\ket{ijkl}$ is labelled as $\ket{ijkl}_{D_A,D_B,\mathrm{Dem}_A,\mathrm{Dem}_B}$.
In our experiments, we always measure the singles rates in modes $D_A$ and $D_B$ to observe the power of the demon.
From Eq. \ref{eq:fullstate}, we can extract the probability for a detector in mode $D_A$ or mode $D_B$ to click.
\begin{eqnarray}\nonumber
    P_A=2P(0,0)P(1,1) + 2P(1,0)P(1,1)  + 
    P(0,0)P(1,0) + P(0,1)P(1,1) +P^2(1,0) + P^2(1,1)\\ 
    \\
    P_B=P(0,0)P(1,0) + 2P(1,0)P(0,1)  +
    2P(1,0)P(1,1) +P(0,1)P(1,1)   
    + P^2(1,0)+P^2(1,1).\nonumber \\
\end{eqnarray}

Now experimentally, we run this setup many times, observing average singles rates at both detectors.
In general, these rates will be given by $\Gamma P_A$ and $\Gamma P_B$, where $\Gamma $ is the total number of repetitions of the experiment.
Thus the measured photon number difference is given by $\Delta N = \Gamma (P_A-P_B)$.
Finally, using our expressions for $P_A$, $P_B$  we arrive at:
\begin{eqnarray}\nonumber
    \Delta N&=&(2P(0,0)P(1,1)-2P(0,1)P(1,0))\Gamma \\
    &=&\frac{2\overline{n}^2}{(1-\overline{n})^4}R^2(1-R^2)\Gamma,
\end{eqnarray}
{where we have substituted $P(n,m)$ from Eq. \ref{eq:pnmap} to arrive at the last line.}
As a last step, we construct our figure of merit for the power of the MD $\nicefrac{\Delta N}{N}$. Here $N$ is the number of input photons in mode $\mathrm{In_A}$, which we constrain to be equal to the number of input photons in mode $\mathrm{In_B}$.
This figure or merit can be interpreted as the demon power per photon.
In the low-photon limit, the probability for one photon to be in mode $\mathrm{In_A}$ is given by $P(1)=\nicefrac{\overline{n}}{(1+\overline{n})}$.
Then the number of single-photon events detected per second is $N=\Gamma P(1)$, and thus
\begin{equation}
    \frac{\Delta N}{N}=\frac{2\overline{n}}{(1-\overline{n})^2}R^2(1-R^2).
\end{equation}
Thus we see that, even in the low photon number limit, the demon effect depends on the effective temperature of the thermal state via $\overline{n}$. As we will see in the next sections, this is not the case for correlated and anti-correlated states in this regime.

\subsection{Split Thermal States}

In this case, the input to the experiment is a single thermal state, which has been split at a beamsplitter.
Thus, a thermal state is still input into both mode $A$ and $B$; however, correlations will now exist between the two modes.  Roughly speaking, now when the demon detects a photon at (say) detector $\mathrm{Dem}_A$, it can no longer know if the  bunching effect will be stronger in mode $A$ or $B$.

In more detail, the state in modes $\mathrm{In_A}$ and $\mathrm{In_B}$ is now:
\begin{equation}\label{eq:rhosplitap}
    \rho_\mathrm{split}=\sum_{n,m=0}^\infty P(n,m)\ket{n,m}\bra{n,m}_{\mathrm{In_A},\mathrm{In_B}},
\end{equation}
where $P(n,m)$ is defined above, with $R=\nicefrac{1}{\sqrt{2}}$.
As above, we will keep only up to the two-photon terms and consider only the diagonal elements:
\begin{eqnarray}\nonumber 
    \rho_\mathrm{split}\approx&&P(0,0)\ket{0,0}\bra{0,0}_{\mathrm{In_A},\mathrm{In_B}} +P(0,1)\ket{0,1}\bra{0,1}_{\mathrm{In_A},\mathrm{In_B}} + P(1,0)\ket{1,0}\bra{1,0}_{\mathrm{In_A},\mathrm{In_B}} +\\
    &&P(2,0)\ket{2,0}\bra{2,0}_{\mathrm{In_A},\mathrm{In_B}}
    +P(0,2)\ket{0,2}\bra{0,2}_{\mathrm{In_A},\mathrm{In_B}}+
    P(1,1)\ket{1,1}\bra{1,1}_{\mathrm{In_A},\mathrm{In_B}} +\dots
\end{eqnarray}
Next, we will add in the two beamsplitters. 
The beamsplitters enact the following transformations:
\begin{eqnarray}\nonumber
    \ket{0,1}\bra{0,1}_{\mathrm{In_A},\mathrm{In_B}}&\rightarrow&(1-R^2)\ket{0100}\bra{0100}_{A,B,\mathrm{Dem}_A,\mathrm{Dem}_B}+R^2\ket{0001}\bra{0001}_{A,B,\mathrm{Dem}_A,\mathrm{Dem}_B}\\ \nonumber
    \ket{1,0}\bra{1,0}_{\mathrm{In_A},\mathrm{In_B}} &\rightarrow& (1-R^2)\ket{1000}\bra{1000}_{A,B,\mathrm{Dem}_A,\mathrm{Dem}_B}+R^2\ket{0010}\bra{0010}_{A,B,\mathrm{Dem}_A,\mathrm{Dem}_B}\\ \nonumber
    \ket{1,1}\bra{1,1}_{\mathrm{In_A},\mathrm{In_B}}& \rightarrow& (1-R^2)^2\ket{1100}\bra{1100}_{A,B,\mathrm{Dem}_A,\mathrm{Dem}_B}
    +R^2(1-R^2)\ket{1001}\bra{1001}_{A,B,\mathrm{Dem}_A,\mathrm{Dem}_B}\\\nonumber
    &+&R^2(1-R^2)\ket{0110}\bra{0110}_{A,B,\mathrm{Dem}_A,\mathrm{Dem}_B}
    +R^4\ket{0011}\bra{0011}_{A,B,\mathrm{Dem}_A,\mathrm{Dem}_B}\\ \nonumber
    \ket{0,2}\bra{0,2}_{\mathrm{In_A},\mathrm{In_B}}&\rightarrow& (1-R^2)^2\ket{0200}\bra{0200}_{A,B,\mathrm{Dem}_A,\mathrm{Dem}_B}
    +2R^2(1-R^2)\ket{0101}\bra{0101}_{A,B,\mathrm{Dem}_A,\mathrm{Dem}_B}\\\nonumber
    &+&R^4\ket{0002}\bra{0002}_{A,B,\mathrm{Dem}_A,\mathrm{Dem}_B}\\ \nonumber
    \ket{2,0}\bra{2,0}_{\mathrm{In_A},\mathrm{In_B}}&\rightarrow& (1-R^2)^2\ket{2000}\bra{2000}_{A,B,\mathrm{Dem}_A,\mathrm{Dem}_B}+2R^2(1-R^2)\ket{1010}\bra{1010}_{A,B,\mathrm{Dem}_A,\mathrm{Dem}_B}\\ 
    &+&R^4\ket{0020}\bra{0020}_{A,B,\mathrm{Dem}_A,\mathrm{Dem}_B}.
\end{eqnarray}
Again, here we have not included the off-diagonal terms. This will not affect our final result since we never recombine two modes, i.e. we never interfere modes after splitting them.
The demon will now attempt to increase the photon number at the detector in mode $D_A$ by swapping modes $A$ and $B$ on detecting a photon at detector $\mathrm{Dem}_B$, as in Eq. \ref{eq:demonTherm}. However, it was pointed out in Ref. \cite{Theo_shu_MD} that the result is the same if the demon instead swaps the modes upon detection of a photon at detector $\mathrm{Dem}_A$.
Substituting in these transformations, and using the fact that $P(n,m)=P(m,n)$ when $R=\nicefrac{1}{\sqrt{2}}$ we arrive at the following expression for the output state:
\begin{eqnarray}\label{eq:fullstatesplit}\nonumber
\rho_\mathrm{out}=&
    P(0,0)\ket{0000}\bra{0000} + P(0,1)\left[(1-R^2)(\ket{0100}\bra{0100}+\ket{1000}\bra{1000})\right. + &\\\nonumber 
    &\left.R^2(\ket{0001}\bra{0001}+\ket{0010}\bra{0010})\right]
        +P(1,1)\left[(1-R^2)^2\ket{1100}\bra{1100}\right.+& \\ \nonumber
    &\left. R^2(1-R^2)(\ket{0101}\bra{0101}+\ket{0110}\bra{0110})+R^4\ket{0011}\bra{0011}\right]+& \\ \nonumber
    &P(0,2)\left[(1-R^2)^2(\ket{2000}\bra{2000}+\ket{0200}\bra{0200}) + R^4(\ket{0020}\bra{0020} + \ket{0002}\bra{0002}) \right.&
    \\ 
    &\left.+ 2R^2(1-R^2)(\ket{1010}\bra{1010} 
    + \ket{1001}\bra{1001})\right],&
\end{eqnarray}
where the mode labels are $D_A$, $D_B$, $\mathrm{Dem}_A$, $\mathrm{Dem}_B$.

From here, we compute the probability for detector $D_A$ and $D_B$ to click:
\begin{eqnarray}
    P_A&=&P(0,1)(1-R^2)+P(1,1)(1-R^2)^2+P(0,2)(1-R^2)^2+4P(0,2)R^2(1-R^2)\\
    P_B&=&P(0,1)(1-R^2)+P(1,1)(1-R^2)^2+P(0,2)(1-R^2)^2+2P(1,1)R^2(1-R^2).
\end{eqnarray}
Then, we calculate the photon number difference $\Delta N=\Gamma (P_A-P_B)$.
\begin{eqnarray}
    P_A-P_B=2R^2(1-R^2)(2P(0,2)-P(1,1))=0,
\end{eqnarray}
where $0$ is obtained after substituting in the explicit expressions for $P(0,2)$ and $P(1,1)$ with $R=\nicefrac{1}{\sqrt{2}}$.
Hence,
\begin{equation}
    \frac{\Delta N}{N}=0,
\end{equation}
for all values of the reflectivity of the demon's beamsplitter. 
Moreover, in Ref. \cite{Theo_shu_MD} it was shown that this result holds in general, not just in the low-photon number regime.

\subsection{Correlated Thermal States}\label{app:corrlated}

To generate our correlated thermal states, we use a type-II SPDC source.
In such a source, photons are always emitted in pairs.
However, when the idler (signal) beam is traced out, the idler (signal) beam is left in a thermal state \cite{loudon2000quantum}.
The general input state is given by
\begin{equation}\label{eq:SPDC_general}
    \ket{\psi}=\frac{1}{\cosh{s}}\sum_{n=0}^\infty(\tanh{s})^n\ket{n,n}_{\mathrm{In_A},\mathrm{In_B}},
\end{equation}
but for our purposes we will only consider up to the two photon term
\begin{equation}\label{eq:spdcIDEAL}
    \ket{\psi}\approx \frac{1}{1-\frac{s^2}{2}}\ket{0,0}_{\mathrm{In_A},\mathrm{In_B}}+ \frac{s}{1-\frac{s^2}{2}}\ket{1,1}_{\mathrm{In_A},\mathrm{In_B}}+\dots.
\end{equation}
Here, $s$ is the so-called squeezing parameter related to the brightness of the SPDC source,and thus to the temperature of the thermal states in each mode: $\mathrm{sinh}^2(s)=\overline{n}$.
Since the SPDC source is pumped with a continuous-wave laser, the terms of order higher than 2 ($\ket{2,2} + \ket{3,3}$…) in Eqs.~\eqref{eq:SPDC_general} and~\eqref{eq:spdcIDEAL} have low probability and can be neglected. I.e. we are in the `low photon-number' regime.

For the uncorrelated thermal states, we did not need to consider loss, since a thermal state remains a thermal state upon losing a photon.
However, here, the correlations are essential so loss will play a large roll in the power of the demon (the information need to extract work).
We model loss as a beamsplitter, which takes $\ket{1}_A\rightarrow \epsilon\ket{1,0}_{\mathrm{In_A},l_A}+\sqrt{1-\epsilon^2}\ket{0,1}_{\mathrm{In_A},l_A}$.  In words, the photon remains in the mode with probability $\epsilon^2$.

Assuming the same loss in each mode, we have
\begin{eqnarray}\label{eq:spdcLOSS}\nonumber
    \ket{\psi}&\approx& \frac{1}{1-\frac{s^2}{2}}\ket{0,0,0,0}_{{\mathrm{In_A},\mathrm{In_B}},l_A,l_B}+\frac{s\epsilon^2}{1-\frac{s^2}{2}}\ket{1,1,0,0}_{{\mathrm{In_A},\mathrm{In_B}},l_A,l_B}\\  &&
    +  \frac{s(1-\epsilon^2)}{1 - \frac{1}{s^2}}\ket{0011}_{{\mathrm{In_A},\mathrm{In_B}},l_A,l_B}
    \frac{s\epsilon\sqrt{1-\epsilon^2}}{1-\frac{s^2}{2}}\left(\ket{1,0,0,1}_{{\mathrm{In_A},\mathrm{In_B}},l_A,l_B}
    +\ket{0,1,1,0}_{{\mathrm{In_A},\mathrm{In_B}},l_A,l_B}\right).
\end{eqnarray}
Next, we will add in the action of the demon's beamsplitters, which will implement the following operation on a photon in mode $A$:
$\ket{1,0}_\mathrm{\mathrm{In_A},E_A}\rightarrow\sqrt{1-R^2}\ket{1,0}_\mathrm{A,\mathrm{Dem}_A}+R\ket{0,1}_\mathrm{A,\mathrm{Dem}_A}$, and the same on mode $B$, where $E_i$ labels the empty input mode the beamsplitters.
\begin{eqnarray}\nonumber
    \ket{\psi}&\approx& \left( \frac{1}{1-\frac{s^2}{2}}\ket{0,0,0,0,0,0}\right. + \frac{s(1-\epsilon^2)}{1 - \frac{1}{s^2}}\ket{0,0,0,0,1,1} \frac{s\epsilon\sqrt{1-\epsilon^2}}{1-\frac{s^2}{2}}R\left( \ket{0,0,1,0,0,1}+\ket{0,0,0,1,1,0}\right) \\\nonumber
    &&+\frac{s\epsilon\sqrt{1-\epsilon^2}}{1-\frac{s^2}{2}}\sqrt{1-R^2}\left( \ket{1,0,0,0,0,1}+\ket{0,1,0,0,1,0}\right)+ \\ \nonumber 
     &&\frac{s\epsilon^2}{1-\frac{s^2}{2}}R\sqrt{1-R^2}\left(\ket{0,1,1,0,0,0}+\ket{1,0,0,1,0,0}\right) + \\
    &&\left.\frac{s\epsilon^2}{1-\frac{s^2}{2}}\left((1-R^2)\ket{1,1,0,0,0,0}+R^2\ket{0,0,1,1,0,0}\right) \right),
\end{eqnarray}
where we have dropped the mode labels but they are $\ket{i,j,k,l,m,n}_\mathrm{A,B,\mathrm{Dem}_A,\mathrm{Dem}_A,l_A,l_B}$.
Finally, the demon uses its optical switch when it measures a photon at detector $\mathrm{Dem}_A$, taking $\ket{i,j,1,0,0,0}\rightarrow\ket{j,i,1,0,0,0}$. In the above equation, this only changes one term, $\ket{0,1,1,0,0,0}\rightarrow\ket{1,0,1,0,0,0}$.
This results in the following state:
\begin{eqnarray}\label{eq:fullstatespdc}\nonumber
    \ket{\psi}&\approx& \left( \frac{1}{1-\frac{s^2}{2}}\ket{0,0,0,0,0,0}+\frac{s(1-\epsilon^2)}{1 - \frac{1}{s^2}}\ket{0,0,0,0,1,1}\right. + 
    \frac{s\epsilon\sqrt{1-\epsilon^2}}{1-\frac{s^2}{2}}R\left( \ket{0,0,1,0,0,1}+\ket{0,0,0,1,1,0}\right) \\\nonumber
    &&+\frac{s\epsilon\sqrt{1-\epsilon^2}}{1-\frac{s^2}{2}}\sqrt{1-R^2}\left( \ket{1,0,0,0,0,1}+\ket{0,1,0,0,1,0}\right)+ \\\nonumber
    && \frac{s\epsilon^2}{1-\frac{s^2}{2}}R\sqrt{1-R^2}\left(\ket{1,0,1,0,0,0}+\ket{1,0,0,1,0,0}\right) + \\
    &&\left.\frac{s\epsilon^2}{1-\frac{s^2}{2}}\left((1-R^2)\ket{1,1,0,0,0,0}+R^2\ket{0,0,1,1,0,0}\right) \right),
\end{eqnarray}
where the mode labels are now $\ket{i,j,k,l,m,n}_\mathrm{D_A,D_B,\mathrm{Dem}_A,\mathrm{Dem}_A,l_A,l_B}$.
Finally, from this we can compute the probability for detectors $D_A$ and $D_B$ to click:
\begin{eqnarray}
    P_A&=&\frac{s^2}{\left(1-\frac{s^2}{2}\right)^2}\left( \epsilon^2(1-\epsilon^2)(1-R^2)+\epsilon^4(1-R^2)^2 +2\epsilon^4R^2(1-R^2)\right)\\ 
    P_B&=&\frac{s^2}{\left(1-\frac{s^2}{2}\right)^2}\left( \epsilon^2(1-\epsilon^2)(1-R^2)+\epsilon^4(1-R^2)^2\right).
\end{eqnarray}
Then, as before, our experimentally measured $\Delta N$ is given by $\Gamma (P_A-P_B)$, and thus
\begin{equation}
    \Delta N =\frac{s^2}{\left(1-\frac{s^2}{2}\right)^2}2\epsilon^4R^2(1-R^2)\Gamma .
\end{equation}

Again, our figure of merit is the demon effect per photon.  However, we now have two ways of computing this from our experimental data.
First of all, we can proceed as we did with the uncorrelated thermal light and normalise by the input singles rate.
In this case, the probability for a photon to be incident in mode $\mathrm{In_A}$ (or in mode $\mathrm{In_B}$) is $\nicefrac{s^2}{\left(1-\nicefrac{s^2}{2}\right)^2}\epsilon^2$, coming directly from Eq. \ref{eq:spdcLOSS}.
Thus, the incoming photon rate is $N=\Gamma \nicefrac{s^2}{\left(1-\nicefrac{s^2}{2}\right)^2}\epsilon^2$, and our figure of merit when normalising by singles is 
\begin{equation}\label{eq:spdcSING}
    \frac{\Delta N}{N}=2\epsilon^2R^2(1-R^2).
\end{equation}
After a closer look at Eq. \ref{eq:spdcLOSS}, one can see that $\epsilon^2$ can be understood as the coupling efficiency. In other words, it is the ratio of the measured coincidence rate to the singles rate before the demon acts.
We thus see that in this case the demon effect is scaled down by the coupling efficiency, which makes sense since a lower coupling efficiency degrades the correlations, making the demon less effective.
The other point to note is that, in this regime, the demon is no longer driven by the thermal statistics of the light, but the correlations from the SPDC dominate.
This is even more apparent when we instead compute the demon effect per photon pair.

Examining Eq. \ref{eq:spdcLOSS} again, we see that the probability for a correlated pair to be incident in mode $\mathrm{In_A}$ and $\mathrm{In_B}$ is given by $P_{AB}=\nicefrac{s^2}{\left(1-\nicefrac{s^2}{2}\right)^2}\epsilon^4$, and thus the coincidence rate is $C=\Gamma P_{AB}$, so
\begin{equation}\label{eq:spdcCOIN}
    \frac{\Delta N}{C}=2R^2(1-R^2).
\end{equation}
Now the coupling efficiency drops of altogether, and we are left with the ideal demon effect for correlated thermal states.
We also see that when the demon uses a 50:50 beamsplitter ($R=\nicefrac{1}{\sqrt{2}}$), then $\nicefrac{\Delta N}{C}=0.5$.
This means that on average, the demon is able to create a photon number imbalance of $0.5$ per photon pair.

\subsection{Anti-Correlated states}
In Ref. \cite{Theo_shu_MD}, a specific two-mode state was proposed with two important properties. 
First, for every N-photon event, the two modes are in a N00N state. 
Second, tracing out either mode, leaves the remaining mode in a thermal state.
Here we do not create this state exactly, but we do note that some experimentally-realised N00N state sources may come close to these conditions \cite{afek2010high,rozema2014scalable}, as they do result in a superposition of N00N states for various photon numbers.
Instead, here we perform Hong-Ou-Mandel interference between the two photons from our  our SPDC source to create a 2-photon N00N state.  The higher-order modes will not be in a N00N state; rather, they will be in so-called Holland-Burnett states \cite{holland1993interferometric}.
The main qualitative difference between these Holland-Burnett states and the anti-correlated state of Ref. \cite{Theo_shu_MD} is that the Holland-Burnett states contain only terms with even photon numbers, while those of Ref. \cite{Theo_shu_MD} contain all photon numbers.
Since experimentally we are in the low-photon number regime (where the probability to detect two photons is much larger than the probability to detect four photons) the main difference between our experimentally-generated state and that of Ref. \cite{Theo_shu_MD} is the absence of the one-photon component.
It is straight-forward to see that this term has no effect on the MD's power. The one-photon term is:
 $\frac{1}{\sqrt{2}}\ket{1,0}_{\mathrm{In_A},\mathrm{In_B}+}\frac{1}{\sqrt{2}}\ket{0,1}_{\mathrm{In_A},\mathrm{In_B}}$.
After application of the demon's beamsplitters, the state is:
\begin{eqnarray}\nonumber
&\frac{1}{\sqrt{2}}\left[ \sqrt{1-R^2}\ket{1,0,0,0}_{A,B,\mathrm{Dem}_A,\mathrm{Dem}_B} +  R\ket{0,0,1,0}_{A,B,\mathrm{Dem}_A,\mathrm{Dem}_B}\right. \\ &+ \left.\sqrt{1-R^2}\ket{0,1,0,0}_{A,B,\mathrm{Dem}_A,\mathrm{Dem}_B} R\ket{0,0,0,1}_{A,B,\mathrm{Dem}_A,\mathrm{Dem}_B}\right].
\end{eqnarray}
Clearly, this state does not change regardless of the demon's action with the UFOS upon detection at $\mathrm{Dem}_A$ or $\mathrm{Dem}_B$.
Thus, the probability to detect a photon at $D_A$ or $D_B$ is equal; i.e. $P_A=P_B=\nicefrac{1-R^2}{2}$. Since $\Delta N$ is proportional to $P_A-P_B$, $\Delta N=0$ for this state. Therefore, the absence of the one-photon component will not affect our measurement of $\Delta N$. 

To model our experimentally produced state, we begin with
\begin{eqnarray}
    \ket{\psi}&=&\frac{1}{1-\frac{s^2}{2}}\ket{0,0}_{\mathrm{In_A},\mathrm{In_B}}+\frac{s}{1-\frac{s^2}{2}}\left( \frac{v}{\sqrt{2}}(\ket{2,0}_{\mathrm{In_A},\mathrm{In_B}} +\ket{0,2}_{\mathrm{In_A},\mathrm{In_B}})+ \sqrt{1-v^2}\ket{1,1}_{\mathrm{In_A},\mathrm{In_B}} \right).\label{eq:noonIdeal}
\end{eqnarray}
Here, again, $s$ is the squeezing parameter.
We have also modeled the imperfect HOM visibility by the parameter $v$, whereby the photons bunch into a N00N state with probability $v^2$.

We then proceed as before, modeling loss as $\ket{1}_\mathrm{In_A}\rightarrow \epsilon\ket{1,0}_\mathrm{\mathrm{In_A},l_A} + \sqrt{1-\epsilon^2}\ket{0,1}_\mathrm{\mathrm{In_A},l_A}$, which acts as $\ket{2}_\mathrm{\mathrm{In_A}}\rightarrow \epsilon^2\ket{2,0}_\mathrm{\mathrm{In_A},l_A} + (1-\epsilon^2)\ket{0,2}_\mathrm{\mathrm{In_A},l_A} + \sqrt{2}\epsilon\sqrt{1-\epsilon^2}\ket{1,1}_\mathrm{\mathrm{In_A},l_A}$ on the two-photon terms.
With the same definitions for mode $B$.
Similarly, the demon's beamsplitters enact the following transformations: $\ket{1,0}_\mathrm{\mathrm{In_A},E_A}\rightarrow \sqrt{1-R^2}\ket{1,0}_\mathrm{A,\mathrm{Dem}_A} + R\ket{0,1}_\mathrm{A,\mathrm{Dem}_A}$, and $\ket{2,0}_\mathrm{\mathrm{In_A},E_A}\rightarrow (1-R^2)\ket{2,0}_\mathrm{A,\mathrm{Dem}_A} + R^2\ket{0,2}_\mathrm{A,\mathrm{Dem}_A} + \sqrt{2}R\sqrt{1-R^2}\ket{1,1}_\mathrm{A,\mathrm{Dem}_A}$.
In this case, the demon's strategy will be the same as for uncorrelated thermal states. This is because when the demon detects a photon in mode $\mathrm{Dem}_B$, it is now more likely that there are more photons in mode $B$ (unless both photons of the N00N state are reflected). Thus the demon will swap the modes.
On the other hand, the demon will not do anything when a photon is detected in mode $\mathrm{Dem}_A$.
Working through all of these transformations, one arrives at the following state:
\begin{eqnarray}\label{eq:fullstatenoon}\nonumber
    \ket{\psi} &=&\frac{1}{1-\frac{s^2}{2}}\ket{0,0,0,0,0,0} + \frac{s}{1-\frac{s^2}{2}}\sqrt{1-v^2}\Big[\epsilon^2(1-R^2) \ket{1,1,0,0,0,0} + \epsilon^2R^2 \ket{0,0,1,1,0,0}  \\ \nonumber
    && +\epsilon^2R\sqrt{1-R^2}(\ket{0,1,1,0,0,0}+\ket{0,1,0,1,0,0})
    \\ \nonumber
    && +\epsilon\sqrt{1-\epsilon^2}\sqrt{1-R^2}(\ket{1,0,0,0,0,1}+\ket{0,1,0,0,1,0}) \\ \nonumber
    &&+ \epsilon\sqrt{1-\epsilon^2}R(\ket{0,0,1,0,0,1}+\ket{0,0,0,1,1,0}) + (1-\epsilon^2)\ket{0,0,0,0,1,1} \Big]  \\ \nonumber
    && +\frac{s}{1-\frac{s^2}{2}}\frac{v}{\sqrt{2}}\Big[\epsilon^2(1-R^2) (\ket{2,0,0,0,0,0}+\ket{0,2,0,0,0,0}) + \epsilon^2R^2 (\ket{0,0,2,0,0,0}+\ket{0,0,0,2,0,0})  \\ \nonumber
&& +\sqrt{2}\epsilon^2R\sqrt{1-R^2}(\ket{1,0,1,0,0,0}+\ket{1,0,0,1,0,0})
\\ \nonumber
&&
+\sqrt{2}\epsilon\sqrt{1-\epsilon^2}\sqrt{1-R^2}(\ket{1,0,0,0,1,0}+\ket{0,1,0,0,0,1})  \\ \nonumber
    && +\sqrt{2}\epsilon\sqrt{1-\epsilon^2}R(\ket{0,0,1,0,1,0}+\ket{0,0,0,1,0,1}) + (1-\epsilon^2)(\ket{0,0,0,0,2,0}+\ket{0,0,0,0,0,2}) \Big].\\
\end{eqnarray}
The mode labels are $\ket{i,j,k,l,m,n}_\mathrm{D_A,D_B,\mathrm{Dem}_A,\mathrm{Dem}_B,l_A,l_B}$.  It is then straightforward to compute the probability for the detectors in mode $A$ and $B$ to click then. For that we can obtain the photon number difference as before
\begin{equation}
    \Delta N=\frac{s^2}{\left(1-\frac{s^2}{2}\right)^2}2(2v^2-1)\epsilon^4R^2(1-R^2){\Gamma }.
\end{equation}
This expression is very similar to that of the SPDC light, simply with an additional factor of $(2v^2-1)$.
So we see in the low photon-number regime, for an ideal source both the correlated and anti-correlated states behave identically.
But in practice, the correlated state should perform better, since they do not require the additional experimental step of Hong-Ou-Mandel 

As a last step, we again compute our two figures of merit, by normalising $\Delta N$ by both the singles and the coincidence counts, yielding
\begin{equation}\label{eq:noonSING}
    \frac{\Delta N}{N}=2\epsilon^2(2v^2-1)R^2(1-R^2)
\end{equation}
and
\begin{equation}\label{eq:noonCOIN}
    \frac{\Delta N}{C}=2(2v^2-1)R^2(1-R^2).
\end{equation}
Similar to the correlated states, we see that normalising by coincidences removes the effect of the imperfect coupling efficiency.
However, the visibility factor remains.

\section{The Maxwell's Demon's Information}
\label{app:info}
In order to quantify the information accessible to the MD we compute the mutual information between the MD's measurement outcomes (detector clicks at detectors $\mathrm{Dem}_A$ and $\mathrm{Dem}_B$) and the photon numbers distribution in mode $A$ and $B$ after MD's two beamsplitters.
As discussed in the main text, this quantity imposes an upper limit on the amount of work that the MD can extract.
The mutual information between the detector outcomes and the photon number quantifies the common information between these systems, and, hence, how much the MD can learn about the photon number from its measurements.

Following Ref.~\cite{Barbieri16MD}, the mutual information between the demons measurement clicks ($M_a$ and $M_b$) and the photon number in mode $A$ and $B$ after the beamsplitter ($N_a$ and $N_b$) in our setup can be written as
\begin{equation}
    I(M_a,M_b:N_a,N_b)=\sum_{m_A}\sum_{m_B}\sum_{n_A}\sum_{n_B} p(m_A,m_B,n_A,n_B)\left[ log(p(m_A,m_B|n_A,n_B))-log(p(m_A,m_B)) \right].
\end{equation}\label{eq:fullI}
In Eq.~\eqref{eq:fullI}, $p(m_A,m_B,n_A,n_B)$ is the probability to detect $m_A$ photons at detector $\mathrm{Dem}_A$, $m_B$ photons at detector $\mathrm{Dem}_B$, while having $n_A$ photons remaining in mode $A$ and $n_B$ photons remaining in mode $B$.
Similarly, $p(m_A,m_B)$ is the probability for the demon to detect $m_A$ and $m_B$ photons at detectors $\mathrm{Dem}_A$ and $\mathrm{Dem}_B$, respectively.
We obtain $p(m_A,m_B)$ by marginalizing $p(m_A,m_B,n_A,n_B)$ over $n_A$ and $n_B$.
The joint probability $p(m_A,m_B,n_A,n_B)$ will depend on the input state, and can be taken almost directly from the output states computed in Appendix \ref{app:MDP}.
In particular, they are obtained from the coefficients of the various terms in Eq.~\eqref{eq:fullstate} for the uncorrelated case, Eq.~\eqref{eq:fullstatesplit} for the split thermal state, Eq.~\eqref{eq:fullstatespdc} for the correlated state, and Eq.~\eqref{eq:fullstatenoon} for the anti-correlated state.
However, the MD's operation given in Eq.~\eqref{eq:demonTherm} must first be undone. Furthermore, for the correlated and anti-correlated states, we ignore the vacuum component from the state preparation (i.e. we set $\ket{\psi}=\ket{1,1}_{A,B}$ in Eq.~\eqref{eq:spdcIDEAL}, and similarly in Eq.~\eqref{eq:noonIdeal}) to reflect our normalization by the singles rate.

The remaining quantity in Eq.~\eqref{eq:fullI} is the conditional probability $p(m_A,m_B|n_A,n_B)$ which is the probability that the MD will detect $m_A$ and $m_B$ photons at detectors $\mathrm{Dem}_A$ and $\mathrm{Dem}_B$, respectively, \emph{given} that $n_A$ and $n_B$ photon are incident in modes $\mathrm{\mathrm{In_A}}$ and $\mathrm{In_B}$, respectively. Since the modes do not interact before the MD's beamsplitter this probability factorizes $p(m_A,m_B|n_A,n_B)= p(m_A|n_A)p(m_B|n_B)$.
These individual probabilities for mode $A$ and $B$ can be readily computed by assuming a Fock state $\ket{n}$ is input to a beamsplitter of reflection amplitude $R$
\begin{equation}
p(m|n)=\sum_{k=0}^{n}\frac{n!~\epsilon^{2k}(1-\epsilon^2)^{n-k}R^{2m}(1-R^2)^{k-m}}{m!(n-k)!(k-m)!},
\end{equation}
where the coupling efficiency is $\epsilon^2$ (as defined above).

Notice that the conditional probabilities are the same for every input state, as they are related just to the MD's apparatus.
The effect of the input states on the mutual information arises through the joint probabilities coming from the calculations presented in Appendix~\ref{app:MDP}, and these introduce dependence the average photon number (for the uncorrelated and split thermal states) and the visibility (for the anti-correlated state).
\end{document}